\documentclass[12pt]{article} 

\usepackage[utf8]{inputenc}
\usepackage[T1]{fontenc} 
\usepackage[english]{babel}           

\usepackage[pdftex]{graphicx}
\usepackage{epsfig}
\usepackage{graphicx}
\usepackage{comment}
\usepackage{latexsym}
\usepackage{hyperref}
\usepackage{amsmath}
\usepackage{mathrsfs}
\usepackage[usenames, dvipsnames]{color}
\usepackage{amsbsy}
\usepackage{amssymb}
\usepackage{amsthm}
\usepackage{amsfonts}
\usepackage{dsfont}
\usepackage{cite}
\usepackage{enumitem}
\usepackage{xcolor}
\usepackage{color}
\usepackage{mathtools}
\usepackage{caption} 
\usepackage{subcaption} 
\usepackage[toc,page]{appendix} 

\usepackage{diagbox}

\newcommand{\be}[0]{\begin{equation}}
\newcommand{\ee}[0]{\end{equation}}

\newcommand{\Eq}[1]{Eq.~\eqref{#1}}
\newcommand{\Ref}[1]{Ref.~\cite{#1}}
\newcommand{\Sect}[1]{Sect.~\ref{#1}}

\renewcommand{\thefootnote}{\fnsymbol{footnote}}

\newcommand{\colvec}[2][.8]{\scalebox{#1}{\renewcommand{\arraystretch}{.8}$\begin{bmatrix}#2\end{bmatrix}$}}

\newcommand{\Teich}{Theichm\"uller }

\newcommand{\Z}{\mathbb{Z}}

\renewcommand{\O}{{\cal O}}

\renewcommand{\Im}{{\rm Im}\,}

\newcommand{\tr}{\textrm{tr}\,}

\newcommand{\eg}{{\it e.g.} }
\newcommand{\ie}{{\it i.e.} }
\newcommand{\via}{{\it via} }

\newcommand{\where}{\mbox{where}}

\renewcommand{\and}{\mbox{and}}

\newcommand{\esp}{\phantom{\!\!\overset{\displaystyle |}{.}}}
\newcommand{\espD}{\phantom{\!\!\underset{\displaystyle |}{\cdot}}}

\usepackage{bm}
\usepackage{scalerel}

\newlength\bshft
\def\fakebold#1{\ThisStyle{\ooalign{$\SavedStyle#1$\cr%
  \kern-\bshft$\SavedStyle#1$\cr%
  \kern\bshft$\SavedStyle#1$}}}

\newcommand{\N}{{\cal N}}
\newcommand{\II}{\mathds{1}}
\newcommand{\T}{{\cal T}}
\newcommand{\Spin}{{\text{Spin}(32)/\Z_2}}
\newcommand{\f}{\text{f}}
\newcommand{\rL}{\text{L}}
\newcommand{\rR}{\text{R}}
\newcommand{\half}{{1\over 2}}
\newcommand{\goh}{h}
\newcommand{\gog}{g}

\catcode`\@=11
\def\marginnote#1{}
\newcount\hour
\newcount\minute
\newtoks\amorpm
\hour=\time\divide\hour by60 \minute=\time{\multiply\hour by60
\global\advance\minute by-\hour}
\edef\standardtime{{\ifnum\hour<12 \global\amorpm={am}%
        \else\global\amorpm={pm}\advance\hour by-12 \fi
        \ifnum\hour=0 \hour=12 \fi
        \number\hour:\ifnum\minute<10 0\fi\number\minute\the\amorpm}}
\edef\militarytime{\number\hour:\ifnum\minute<10 0\fi\number\minute}
\def\draftlabel#1{{\@bsphack\if@filesw {\let\thepage\relax
   \xdef\@gtempa{\write\@auxout{\string
      \newlabel{#1}{{\@currentlabel}{\thepage}}}}}\@gtempa
   \if@nobreak \ifvmode\nobreak\fi\fi\fi\@esphack}
        \gdef\@eqnlabel{#1}}
\def\@eqnlabel{}
\def\@vacuum{}
\def\draftmarginnote#1{\marginpar{\raggedright\scriptsize\tt#1}}
\def\draft{\oddsidemargin -.2truein
        \def\@oddfoot{\sl preliminary draft \hfil
        \rm\thepage\hfil\sl\today\quad\militarytime}
        \let\@evenfoot\@oddfoot \overfullrule 3pt
        \let\label=\draftlabel
        \let\marginnote=\draftmarginnote
   \def\@eqnnum{(\theequation)\rlap{\kern\marginparsep\tt\@eqnlabel}%
\global\let\@eqnlabel\@vacuum}  }
\def\thebibliography#1{
\vskip 0.5cm \centerline{\bf \Large References}
\list{
[\arabic{enumi}]}{\settowidth\labelwidth{[#1]}
\leftmargin\labelwidth
\advance\leftmargin\labelsep
\usecounter{enumi}}
\def\newblock{\hskip .11em plus .33em minus .07em}
\sloppy\clubpenalty4000\widowpenalty4000
\sfcode`\.=1000\relax}

\renewcommand{\theequation}{\arabic{section}.\arabic{equation}}
\renewcommand{\section}{\setcounter{equation}{0}\@startsection
{section}{1}{0mm}{-\baselineskip}{0.5\baselineskip} {\normalfont\Large\bfseries}}
\renewcommand{\subsection}{\@startsection
{subsection}{2}{0mm}{-\baselineskip}{0.5\baselineskip} {\normalfont\large\bfseries}}
\renewcommand{\subsubsection}{\@startsection
{subsubsection}{3}{0mm}{-\baselineskip}{0.5\baselineskip}
{\normalfont\normalsize\slshape}}

\begin{document}


\begin{titlepage}
\begin{flushright}
CPHT-RR015.032020, June 2020
\vspace{1cm}
\end{flushright}
\begin{centering}
{\boldmath\bf \Large HETEROTIC ORBIFOLDS, REDUCED RANK AND \\ \vspace{.3cm} $SO(2n+1)$ CHARACTERS  
 }

\vspace{5mm}

 {\bf Herv\'e Partouche and Balthazar de Vaulchier}

 \vspace{3mm}

{CPHT, CNRS, Ecole polytechnique, IP Paris, \\F-91128 Palaiseau, France\\
{\em herve.partouche@polytechnique.edu, \\ balthazar.devaulchier@polytechnique.edu}}

\end{centering}
\vspace{0.7cm}
$~$\\
\centerline{\bf\Large Abstract}\\

\begin{quote}

The moduli space of the maximally supersymmetric heterotic string in $d$-dimen\-sional  Minkowski space contains various components characterized by the rank of the gauge symmetries of the vacua they parametrize. We develop an approach for describing in a unified way continuous Wilson lines which parametrize a component of the moduli space, together with discrete deformations responsible for the switch from one component to the other. Applied to a component that contains vacua with $SO(2n+1)$ gauge-symmetry factors, our approach yields a description of all backgrounds of the component in terms of free-orbifold models. The orbifold generators turn out to act symmetrically or asymmetrically on the internal space, with or without discrete torsion. Our derivations use extensively affine characters of $SO(2n+1)$. As a by-product,  we find a peculiar orbifold description of the heterotic string in ten dimensions, where all gauge degrees of freedom arise as twisted states, while the untwisted sector reduces to the gravitational degrees of freedom. 

\end{quote}

\end{titlepage}
\newpage
\setcounter{footnote}{0}
\renewcommand{\thefootnote}{\arabic{footnote}}
 \setlength{\baselineskip}{.7cm} \setlength{\parskip}{.2cm}

\setcounter{section}{0}


\section{Introduction}

Since the seminal works of \Ref{characters2, characters1}, the use of affine characters of simply-laced Lie groups has become  very common. For instance, in ten dimensions, the characters of $SO(8)$ are used in the construction of the type I, type II and heterotic strings, while those of $SO(32)$ and $E_8$ are employed in the description of the gauge degrees of freedom of the $\Spin$ and $E_8\times E_8$ heterotic strings. On the contrary, the use of characters of non simply-laced Lie groups, such as those of $SO(2n+1)$ introduced in \Ref{odd_characters},  are far more sparse in the string-theory literature.  The main goals of the present work are then to 

$(i)$ present  interesting practical applications of $SO(2n+1)$ characters, such as describing models with gauge groups of reduced ranks,

$(ii)$ stress subtleties in the use of the $SO(2n+1)$ spinorial characters, 

$(iii)$ and develop a systematic way of constructing consistent  models based on $SO(2n+1)$ affine algebras, together with their marginal deformations. 

\noindent All our results are derived in the framework of the $\Spin$ or $E_8\times E_8$ heterotic string theories compactified toroidally. 

One feature of models describing  $SO(2n+1)$ gauge group factors is that the rank of the full gauge symmetry cannot be maximal. The underlying reason of this fact is that if one breaks  $SO(2n)$ into a subgroup $SO(2k_1+1) \times SO(2k_2-1)$ where $k_1+k_2=n$, then the initial rank $n$ is reduced to $n-1$.  Of course, such a breaking cannot be realized spontaneously by switching on moduli in the Coulomb branch. On the contrary, it must be produced by applying some discrete deformation that maps the initial setup into a model belonging to a distinct component of the moduli space of the heterotic string~\cite{triple}. We will first elaborate further on the possibility introduced in \Ref{so3so1,GKP} to implement such deformations \via free orbifold actions. 

Alternatively, we develop another approach which unifies the implementations of continuous Wilson lines and discrete deformations responsible for the reduction of the rank. This point of view is in the spirit of the technical similarity between the implementation of Wilson-lines and the breaking of supersymmetry \`a la Scherk--Schwarz in string theory~\cite{SSstring4,SSstring2,SSstring3,SSstring1,421,CP}. Indeed, such a super-Higgs mechanism amounts to introducing a \emph{discrete} deformation  of the parent supersymmetric model.\footnote{Higgsing in the Coulomb branch and super-Higgs mechanism are both based on underlying  worldsheet symmetries. However, that involved in the super-Higgs case must preserve the worldsheet supercurrent, which imposes a quantization of the deformation parameter.}  

The first example of model involving $SO(2n+1)$ gauge group factors we  consider is an extreme case, in the sense that all of the $SO(32)$ gauge symmetry is broken to a trivial $SO(1)^{32}$, where ``$SO(1)$'' is the group containing only the neutral element.  To construct it, we proceed in two steps. Our starting point in \Sect{d10} is the $\Spin$ theory in ten dimensions, on which we implement  a $\Z_2^5$ orbifold action on the $\Spin$ root lattice whose effect is to project out all gauge bosons from the untwisted sector. However, because there are only two supersymmetric heterotic string theories in ten dimensions~\cite{het-1,het-2}, all 496 gauge bosons must be recovered in the twisted sectors. As a result, we obtain a very peculiar description of the heterotic string where, at the massless level,  the untwisted sector contains only the gravitational multiplet, while all gauge vector multiplets are made of  twisted states. 

In a second step, we compactify in \Sect{orb1616} the above $\Z_2^5$-orbifold setup on a torus $T^5$, while imposing each generator to also act as a half-period translation along a compact direction. As a result, the orbifold group becomes freely acting, ensuring all twisted sectors to become massive. Hence, in the model in five dimensions, the initial $SO(32)$ gauge symmetry is reduced by the $\Z_2^5$ free-orbifold action to nothing~\cite{so3so1,GKP}, or rather $SO(1)^{32}$, as demonstrated by manipulating affine characters. The rank of the gauge-symmetry group  is thus reduced by 16 units. 

We also show in \Sect{orb1616} that imposing only four of the five generators to be free, one of the twisted sectors contains at the massless level 16 Abelian vector multiplets, thus restoring maximal rank. We prove this in various ways, one of which allowing us to stress subtleties concerning the use of the spinorial character of $SO(2n+1)$ in one-loop partition functions. The point is that $SO(2n)$ has two spinorial conjugacy classes of opposite chiralities~\cite{characters2,BLT}, while  $SO(2n+1)$ has only one, which is non-chiral~\cite{odd_characters}. Hence, when decomposing a spinorial character of $SO(2n)$ into spinorial characters of $SO(2k_1+1)$ and $SO(2k_2-1)$ where $k_1+k_2=n$, one is led to omit contributions involving vanishing Jacobi modular forms $\vartheta\colvec[0.7]{1/2\\1/2}$ which capture information on the chirality of the initial  $SO(2n)$ characters. However, such vanishing contributions in partition functions showing up in untwisted (twisted) sectors are related by the action of orbifold-group elements to non trivial contributions in twisted (untwisted) sectors. As a result, when implementing orbifold actions on partition functions, it is sometimes mandatory to keep track of vanishing  contributions $\vartheta\colvec[0.7]{1/2\\1/2}$, even though they have no meaning from the point of view of representation theory of $SO(2k_1+1)$ and $SO(2k_2-1)$.

In \Sect{wl}, we consider the most general lattice involved in toroidal compactification of the $\Spin$ heterotic string in presence of arbitrary Wilson-line background~\cite{421,CP}. We show how this expression can be extended to include discrete deformations that yield a reduction of the rank of the gauge-symmetry group. The backgrounds obtained this way are equivalent to those derived from the free-orbifold point of view, including those acting left/right asymmetrically~\cite{asymmorb} on the coordinates of the internal torus. Moreover, they take into account different choices of discrete torsion~\cite{torsion-1,torsion-2} and keep track of the remaining  marginal deformations. 

Our conclusions can be found in \Sect{cl}. The conventions for the $SO(2n)$ and $SO(2n+1)$ affine characters we use are given in an Appendix, which also lists identities among characters useful for the construction of $\Z_2$-orbifold models. 


\section{Heterotic orbifolds in ten dimensions}
\label{d10}

In order to construct $\N=1$ supersymmetric  heterotic-string theories in ten dimensions, one must  use one of the two even self-dual 16-dimensional Euclidean lattices, which are the root  lattices of $\Spin$ and $E_8\times E_8$~\cite{het-1,het-2}. Moreover, any orbifold action implemented on  a string model in a modular-invariant way yields a consistent theory~\cite{DHVW1,DHVW2}. From the above statements, it follows that any supersymmetry-preserving orbifold action on one of the supersymmetric ten-dimensional heterotic strings must either be trivial, or transform it into the other. In the following, we first review this fact in the simplest cases of $\Z_2$ groups. We then consider orbifold actions that are products of such $\Z_2$'s and show that in some cases gauge and gravitational degrees of freedom may be split in the following sense: All vector multiplets sit in twisted sectors, while the supergravity multiplet arises in the untwisted sector.  

In our conventions, the bosonic side of the heterotic string is holomorphic (right-moving) while the supersymmetric side is antiholomorphic (left-moving). On the holomorphic side, there are 16 extra bosonic coordinates which are compact or, in fermionic language, an equivalent system of 32 real fermions $\psi^i$, $i\in\{1,\dots, 32\}$. On the genus-one surface of \Teich parameter $\tau$, such a fermion has anti-periodic or periodic boundary conditions upon parallel transport along the cycles,
\be
\psi^i(z+1)=-e^{2i\pi \gamma}\psi^i(z)~,\qquad \psi^i(z+\tau)=-e^{2i\pi \delta}\psi^i(z)~,
\ee
where $\gamma=0$ (Neveu--Schwarz) or $\gamma=\half$ (Ramond), and likewise $\delta\in\{0,\half\}$. In ten dimensions, an orbifold action preserving $\N=1$ supersymmetry can only act  upon the $\psi^i$'s. In the present work, we will only consider $\Z_2$ groups that flip $\psi^i\to -\psi^i$, where $i$ belongs to a subset of the 32 fermions. Indeed, such transformations are symmetries of the worldsheet action of the free fields $\psi^i$ and may be used as a basis for constructing orbifold theories. For a fermion sensitive to the $\Z_2$ action, the rules for parallel transport are affected in a way encoded by shifts $\gamma\to \gamma+h$ and  $\delta\to \delta+g$, where $h,g\in\{0,\half\}$.  Hence, its contribution to the one-loop partition function can be expressed in terms of a Jacobi  modular form $\vartheta$  with shifted characteristics, along with a Dedekind function $\eta$,  
\be\label{twist}
\sqrt{\frac{\vartheta\colvec[0.7]{\gamma+h\\\delta+g}(0|\tau)}{\eta(\tau)}}~ .
\ee
In orbifold language, $h=0$ and $h=\half$ correspond respectively to the untwisted and twisted sectors, while summing over $g$ implements the projection onto the $\Z_2$-invariant spectrum.


\subsection{From \boldmath $\Spin$ to $E_8\times E_8$}
\label{h8}

Let us first review the fact that 
{\em a $\Z_2$-orbifold group acting on 16 of the 32 fermions $\psi^i$ transforms the $\mbox{\rm Spin}(32)/\Z_2$ lattice into the $E_8\times E_8$ lattice}~\cite{DHVW2}.\footnote{This can be related to the fact that the  $\Spin$ and $E_8\times E_8$ theories compactified on a circle share the same moduli space parametrized by the radius of the circle and the Wilson-line background~\cite{Ginsparg:1986bx}.} 

Our starting point is the partition function of the 32 fermions of the $\Spin$ heterotic-string theory,  
\be
{\Gamma^\Spin\over \eta^{16}} =\frac{1}{2} \sum_{\gamma,\delta} \left( \frac{\vartheta\colvec[0.7]{\gamma\\\delta}}{\eta} \right)^{16} =O_{32} + S_{32} ~ ,
\label{lat32}
\ee
where $O_{2n}, S_{2n}$ (along with $V_{2n}, C_{2n}$) are the affine characters associated with the four conjugacy classes of $SO(2n)$. Their explicit expressions can be found in \Eq{defchar}. Let us define $G_1$ to be the $\Z_2$-orbifold generator that flips 16 of the 32 fermions $\psi^i$.  Up to a reordering, we may choose the signs of its eigenvalues $\pm 1$ to be respectively   
\be
G_1 : ++++++++++++++++---------------- ~ .
\ee
Our goal is to derive the counterpart of the $\Spin$ lattice in the orbifold theory. The lattice we are looking  for and denote as $\Gamma^\Spin_{G_1}$ may be decomposed into untwisted and twisted contributions, on which projections onto $\Z_2$-invariant weight vectors are applied. This yields four contributions labelled by $h_1,g_1\in\{0,\half\}$, 
\be
\Gamma^\Spin_{G_1}\equiv \half\sum_{h_1,g_1} \Gamma^\Spin\colvec[0.7]{h_1\\g_1}\; .
\label{la1}
\ee
To derive them, we use the fact that $SO(32)\supset SO(16)\times SO(16)$, which implies that the $SO(32)$ characters appearing in \Eq{lat32} may be decomposed in terms of $SO(16)$ characters. Indeed, this can be done explicitly using the identities~(\ref{decomp1}),\footnote{Throughout our work, it is understood that in a monomial that is composed of a product of characters, the latter should not be commuted in order to keep track of which fermions $\psi^i$ they are referring to.} which yield 
\be
O_{32}=O_{16}O_{16}+V_{16}V_{16}~,~~\quad S_{32}=S_{16}S_{16}+C_{16}C_{16}~.
\label{dOS}
\ee

In the untwisted sector,  the partition functions of the 16 fermions flipped by $G_1$ have shifted characteristics  $\colvec[0.7]{\gamma\\\delta+g_1}$. Therefore, we see from their definitions that all $O_{16}$ and $S_{16}$ characters are invariant under the action of $G_1$, while the characters $V_{16}$ and $C_{16}$ that are in second positions in the monomials appearing in \Eq{dOS} acquire signs $(-1)^{2g_1}$. In fact, for arbitrary $SO$-group characters, one may  apply the general formulas~(\ref{Gtr}) obtained by flipping the boundary conditions along the worldsheet cycle $[0,\tau]$.   Hence, we obtain 
\be\label{untwso32}
\frac{\Gamma^\Spin \colvec[0.7]{0\\g_1}}{\eta^{16}} = O_{16}O_ {16} + V_{16}(-1)^{2g_1}V_{16} + S_{16}S_{16} + C_{16}(-1)^{2g_1}C_{16} ~ .
\ee
In the twisted sector,  the partition functions of the  flipped fermions have characteristics $\colvec[0.7]{\gamma+1/2\\ \delta+g_1}$. Hence, all $SO(16)$ characters in second positions in the monomials appearing in \Eq{dOS}  are permuted  according to $O_{16}\leftrightarrow S_{16}$ and $V_{16}\leftrightarrow C_{16}$. In general, the twisted characters for arbitrary $SO$ groups can be derived from untwisted ones by applying the  rules listed in  \Eq{Gtwist}. As a result, we have 
\be\label{twso32}
\frac{\Gamma^\Spin\colvec[0.7]{1/2\\g_1}}{\eta^{16}} = O_{16}S_ {16} + V_{16}(-1)^{2g_1}C_{16} + S_{16}O_{16} + C_{16}(-1)^{2g_1}V_{16} ~ .
\ee
Summing over all contributions as shown in \Eq{la1}, we obtain
\be\label{nfso32}
{\Gamma^\Spin_{G_1}  \over \eta^{16}}= (O_{16}+S_{16})^2 \equiv  \frac{\Gamma^{E_8\times E_8}}{\eta^{16}} ~ ,
\ee
where the last equality, which  involves the root lattice of $E_8\times E_8$, holds thanks to the identity satisfied by the root lattice of $E_8$, 
\be
{\Gamma^{E_8}\over \eta^8} = \half\sum_{\gamma,\delta} \left( \frac{\vartheta\colvec[0.7]{\gamma\\\delta}}{\eta} \right)^8 = O_{16}+S_{16} ~ .
\label{e8}
\ee
Therefore, acting on the $\Spin$ heterotic string with the orbifold group generated by $G_1$  leads to the $E_8\times E_8$ theory.


\subsection{From \boldmath $E_8\times E_8$  to $\Spin$}
\label{8h}

Next, we review the fact that a similar statement exists, where the roles of the two $\N=1$ heterotic theories in ten dimensions is reversed: 
{\em In the $E_8\times E_8$ heterotic string, a $\Z_2$-orbifold group acting on 8 of the 16 fermions $\psi^i$ that generate the first $E_8$ lattice, and also acting on 8 of the remaining fermions $\psi^i$ transforms the $E_8\times E_8$ lattice into the $\mbox{\rm Spin}(32)/\Z_2$ one}~\cite{DHVW2}.

To begin with, we consider the partition function of the $16+16$ fermions $\psi^i$, with appropriate boundary conditions to generate the $E_8\times E_8$ affine Lie algebra, \Eq{nfso32} and~\eqref{e8}. Let $G_2$ be a $\Z_2$-orbifold  generator that satisfies the assumptions stated above. Up to a reordering, its eigenvalues $\pm1$ have signs 
\be
G_2 : ++++++++--------++++++++-------- ~ .
\label{g2}
\ee
Our aim is to derive the lattice appearing in the $E_8\times E_8$  theory orbifolded by $\Z_2$ generated by $G_2$. As before, we may split this lattice  into four pieces, 
\be
\Gamma^{E_8\times E_8}_{G_2} \equiv \half\sum_{h_2,g_2}\Gamma^{E_8\times E_8}\colvec[0.7]{h_2\\g_2}\;,
\ee
where $h_2,g_2\in\{0,\half\}$.
In order to find how $G_2$ transforms the $SO(16)$ characters in \Eq{nfso32}, it is useful to express them in terms of characters of $SO(8)$ using \Eq{decomp1}.
 
In the untwisted sector, the effect of $G_2$ is to multiply by $-1$ all characters $V_8$ and $C_8$ associated with the fermions sensitive to the orbifold, which yields 
\be
\begin{aligned}
{\Gamma^{E_8\times E_8}\colvec[0.7]{0\\g_2}\over \eta^{16}} =&\left( O_8O_8+V_8(-1)^{2g_2}V_8+S_8S_8+C_8(-1)^{2g_2}C_8\right)^2 ~.
\end{aligned}
\ee
Upon summing over $g_2$, the untwisted-sector contribution is therefore  
\be\label{untwe8e8}
\begin{aligned}
\half\sum_{g_2}{\Gamma^{E_8\times E_8}\colvec[0.7]{0\\g_2}\over \eta^{16}} &=O_8^4 + V_8^4 + S_8^4  + C_8^4 + O_8^2S_8^2  + S_8^2O_8^2 + V_8^2C_8^2+ C_8^2V_8^2 \\
&=(O_8^2+S_8^2)^2+(V_8^2+C_8^2)^2= O'_{16}O'_{16} + S'_{16}S'_{16} ~ ,
\end{aligned}
\ee
where the last equality holds thanks to the triality symmetry among  the $SO(8)$ characters. The latter amounts to exchanging $V_8\leftrightarrow S_8$ (in arbitrary positions in the monomials).\footnote{\label{NB} Alternatively, one can permute $S_8\to V_8\to C_8\to S_8$.} In \Eq{untwe8e8}, we use ``primed'' characters in the final expression to keep track of this manipulation. 

In the twisted sector, all $SO(8)$ characters associated with fermions whose boundary conditions are sensitive to the action of $G_2$ are permuted  according to $O_{8}\leftrightarrow S_{8}$ and $V_{8}\leftrightarrow C_{8}$, as follows from \Eq{Gtwist}. Hence, we obtain 
\be
\begin{aligned}
{\Gamma^{E_8\times E_8}\colvec[0.7]{1/2\\g_2}\over \eta^{16}} &=\left( O_8S_8+V_8(-1)^{2g_2}C_8+S_8O_8+C_8(-1)^{2g_2}V_8\right)^2 ~,
\end{aligned}
\ee
which leads to the twisted-sector contribution 
\be
\begin{aligned}
\half\sum_{g_2}{\Gamma^{E_8\times E_8}\colvec[0.7]{1/2\\g_2}\over \eta^{16}} & = (O_8S_8)^2  + (V_8C_8)^2 + (S_8O_8)^2 + (C_8V_8)^2 \\ 
&  ~~~+ O_8S_8S_8O_8 + V_8C_8C_8V_8 + S_8O_8O_8S_8 + C_8V_8V_8C_8 \\
&=(O_8S_8+S_8O_8)^2+(V_8C_8+C_8V_8)^2 = V_{16}'V'_{16} + C'_{16}C'_{16} ~ .
\end{aligned}
\ee
Again, the last equality is found by applying the $SO(8)$-triality symmetry already used in the derivation of \Eq{untwe8e8}, $V_8\leftrightarrow S_8$.$^{\ref{NB}}$

Adding together the contributions of both sectors, we obtain
\be
{\Gamma^{E_8\times E_8}_{G_2}\over \eta^{16}}  = O_{32} + S_{32} = {\Gamma^\Spin \over \eta^{16}} ~ ,
\ee
which means that the $E_8\times E_8$ heterotic string  orbifolded by $\Z_2$ generated by $G_2$  is nothing but the $\Spin$ theory. 


\subsection{Twisted/untwisted descriptions of the spectrum}
\label{23}

As illustrated by the $\Z_2$ actions considered in the previous subsections, orbifolding the heterotic string theories in 10 dimensions while preserving $\N=1$ supersymmetry yields only alternative descriptions. 
In particular, it is a matter of convention to describe (at least part of) the gauge degrees of freedom in twisted or untwisted sectors. For instance, modding out the $\Spin$ string theory by $\Z_2\times \Z_2$ generated by $G_1$ and $G_2$ leads to a framework where the gauge degrees of freedom are realized in one untwisted and three twisted sectors. In the following we show that an extreme case exists for a $\Z_2^5$-orbifold action where, at the massless level, the untwisted sector contains only the $\N=1$ gravitational multiplet, while all 496 non-Abelian $\N=1$ vector multiplets are realized in the $2^5-1=31$ twisted sectors. To reach this conclusion, we are going to see that no gauge degree of freedom  in the  untwisted sector survives the projection onto the states invariant under the generators of the $\Z_2^5$ group.

Let us consider the $\Spin$ theory and define besides $G_1$ and $G_2$   three more $\Z_2$ generators acting on 16 of the 32 fermions $\psi^i$. Altogether, the signs of the  eigenvalues of the generators are given by 
\be
\begin{aligned}
G_1 :&\ ++++++++++++++++---------------- ~ , \\
G_2 :&\ ++++++++--------++++++++-------- ~ , \\
G_3 :&\ ++++----++++----++++----++++---- ~ , \\
G_4 :&\ ++--++--++--++--++--++--++--++-- ~ , \\
G_5 :&\ +-+-+-+-+-+-+-+-+-+-+-+-+-+-+-+- ~ .
\end{aligned}
\ee 
We know that the $\Z_2^5$-orbifold model is either the $E_8\times E_8$ or $\Spin$ theory.\footnote{From the analyses of Sect.~\ref{h8} and \ref{8h}, one may think that the actions of $N$ generators of this type yields the $E_8\times E_8$ theory when $N$ is odd and the $\Spin$ theory when $N$ is even. However, this is not an obvious fact. One can always write down the low lying states of a $\Z_2^N$-orbifold theory to derive their associated Dynkin diagram and figure out whether they realize the $SO(32)$ or $E_8\times E_8$ gauge group.} Splitting the partition function of the 32 fermions $\psi^i$  into $32^2$ pieces, we write
\be
\Gamma^\Spin_{G_1,\dots,G_5}\equiv {1\over 2^5}\sum_{\substack{h_1,\dots,h_5 \\ g_1,\dots, g_5}} \Gamma^\Spin\colvec[0.7]{h_1,\dots,h_5\\g_1,\dots,g_5}\; .
\label{la5}
\ee
Our aim is to show that the untwisted contribution 
\be
{1\over 2^5}\sum_{g_1,\dots,g_5} \Gamma^\Spin\colvec[0.7]{\!\!0\,\,,\dots,0\\g_1,\dots,g_5}
\ee
does not yield any massless state in the full theory.  The key point is  that there is no pair of fermions with identical boundary conditions for all $g_1,\dots, g_5$. 


\paragraph{\em Projection of the untwisted sector:} 

We already know from \Eq{untwso32} that the weight lattice of $\Spin$ projected on \mbox{$G_1$-invariant} states satisfies 
\be
\half\sum_{g_1}\frac{\Gamma^{\Spin}\colvec[0.7]{0 \\ g_1}}{\eta^{16}} =O_{16}^2 + \text{massive} ~ .
\ee
In this equation, the monomial $S_{16}^2$ is implicit in the ``massive'' contribution, as it does not lead to any massless mode (see the $q$-expansion of $S_{16}$ given in \Eq{defchar}, where $q=e^{2i\pi\tau}$).
In order to implement the projection onto $G_2$-invariant states,  we use the identity  $O_{16}^2=(O_8^2+V_8^2)^2$ and obtain immediately 
\be\label{g1g2}
\begin{aligned}
\frac{1}{2^2}\sum_{g_1,g_2}\frac{\Gamma^{\Spin}\colvec[0.7]{0,0 \\ g_1,g_2}}{\eta^{16}} &=\half \sum_{g_2}\big( O_8^2 + V_8(-1)^{2g_2}V_8\big)^2 + \text{massive} \\
& =O_8^4+\text{massive}~,
\end{aligned}
\ee
where we have  included the monomial $V_8^4$ in the ``massive'' contributions (see \Eq{defchar}). Proceeding the same way with the generator $G_3$, we write  $O_8^4=(O_4^2+V_4^2)^4$, which yields
\be
\begin{aligned}\label{g1g2g3}
\frac{1}{2^3}\sum_{g_1,g_2,g_3}\frac{\Gamma^{\Spin}\colvec[0.7]{0,0,0 \\ g_1,g_2,g_3}}{\eta^{16}} &=\half\sum_{g_3}\big( O_4^2 + V_4(-1)^{2g_3}V_4\big)^4 +\text{massive} \\
&=O_4^8 + \text{massive} ~ ,
\end{aligned}
\ee
while for $G_4$ we use $O_4^8=(O_2^2+V_2^2)^8$ and obtain 
\be
\begin{aligned}\label{g1g2g3-2}
\frac{1}{2^4}\sum_{g_1,\dots,g_4}\frac{\Gamma^{\Spin}\colvec[0.7]{0,\dots,0 \\ g_1,\dots,g_4}}{\eta^{16}} &=\half\sum_{g_4}\big( O_2^2 + V_2(-1)^{2g_4}V_2\big)^8 +\text{massive} \\
&=O_2^{16} + \text{massive} ~ .
\end{aligned}
\ee
Notice that  the decompositions of $O_n$ and $V_n$ in~\Eq{decomp1}, the transformation rules of these characters in \Eq{Gtr}, and the $q$-expansion of $V_n$ in \Eq{defchar} hold whatever the parities of $n$ and $p$ in these formulas. Hence, we may implement the last projection onto $G_5$-invariant states by writing 
$O_2^{16}=(O_1^2+V_1^2)^{16}$, which leads to the final result
\be
\begin{aligned}\label{g1g2g3-3}
\frac{1}{2^5}\sum_{g_1,\dots,g_5}\frac{\Gamma^{\Spin}\colvec[0.7]{0,\dots,0 \\ g_1,\dots,g_5}}{\eta^{16}} &=\half\sum_{g_5}\big( O_1^2 + V_1(-1)^{2g_5}V_1\big)^{16} +\text{massive} \\
&=O_1^{32} + \text{massive} ~ .
\end{aligned}
\ee

From the $q$-expansion of the character $O_1$, we see that $O_1^{32}$ yields a non-physical (\ie non-level-matched) tachyonic mode and, more important for us, no state at the massless level. Hence, the untwisted sector contributes an $SO(1)^{32}$ subgroup  of the full gauge symmetry, where $SO(1)$ is the trivial zero-dimensional group, \ie containing only the neutral element. This shows that all gauge degrees of freedom (including the Cartan subgroup) of the \mbox{$\Z_2^5$-orbifold} description of the heterotic string arise from the remaining 31 sectors, which are twisted. Hence, the untwisted sector contains only the gravitational sector. At the massless level, the counting of states goes as follows: In the untwisted sector, there are $8\times 8$ boson/fermions pairs of  degrees of freedom corresponding to the dilaton, graviton and antisymmetric tensor along with their superpartners. In the 31 twisted sectors, there are $8\times 496$ pairs of degrees of freedom associated with non-Abelian vector bosons and their fermionic partners. 


\paragraph{\em Projection of the untwisted Fock space:} 

It is instructive to recover these conclusions by implementing explicitly a projection onto the $\Z_2^5$-invariant states of the untwisted sector of the Fock space. The gauge bosons of the $\Spin$ theory transform in the adjoint representation of $SO(32)$.  They are realized by acting with the low-lying creation operators on the left and right Neveu--Schwarz vacua.\footnote{The right-moving Ramond sector associated with the character $S_{32}$ is massive (see~\Eq{defchar}).} Denoting $\tilde \psi^\mu$ the worldsheet superpartners of the left-moving spacetime coordinates $X_\rL^\mu$, they are  the $(32\times31)/2$ vectors 
\be
 |\mu;i,j\rangle = \tilde\psi_{-\frac{1}{2}}^\mu |\text{NS}\rangle_\text{L} \otimes \psi_{-\frac{1}{2}}^i\psi_{-\frac{1}{2}}^j |\text{NS},h_1=\cdots=h_5=0\rangle_\text{R} ~ ,\quad~~ i<j\in\{1,\dots,32\}~.
\ee
Notice that the latter include those associated with the roots of $SO(32)$ as well as those generating the Cartan subalgebra, which correspond respectively to the massless states arising from the lattice $\Gamma^\Spin$ and the factor $1/\eta^{16}$ in the partition function of the fermions $\psi^i$.

Because $G_1$ flips $\psi_{-\frac{1}{2}}^i\rightarrow-\psi_{-\frac{1}{2}}^i$ for $i\in\lbrace17,...,32\rbrace$, the modes invariant under $G_1$ satisfy  $i,j\in\lbrace1,...,16\rbrace$ or $i,j\in\lbrace17,...,32\rbrace$, and realize the $2\times (16\times 15)/2$ generators of $SO(16)^2$. Applying similar projections onto $G_2$-, $G_3$- and $G_4$-invariant states, the surviving  vectors are $|\mu; 2i-1,2i\rangle$, $i\in\lbrace1,...,16\rbrace$, which generate $SO(2)^{16}$. Hence, all massless states associated with the roots of $SO(32)$ have been projected out and we are left with the Cartan generators. However, because  the action of $G_5$ flips  $\psi_{-\frac{1}{2}}^{2i}\rightarrow-\psi_{-\frac{1}{2}}^{2i}$, $i\in\lbrace1,...,16\rbrace$, the Cartan generators are also projected out. Hence, no gauge boson is realized in the untwisted sector of the $\Z_2^5$ orbifold theory, implying all of the 496 gauge symmetry generators to arise in the 31 twisted sectors. 


\section{Free orbifolds in lower dimensions}
\label{orb1616}

In order to preserve $\N=1$ supersymmetry in ten dimensions, the orbifold groups we have considered so far act on the $\Spin$ or $E_8\times E_8$ lattices only.  In lower dimension~$d$, though, new possibilities can be considered since orbifold groups may also act on the internal space. In this section and the following, we use the fact that free versions of orbifold generators imply twisted sectors to become massive, and can therefore reduce the dimension of the gauge symmetry. More specifically, we show that when vector bosons in the Cartan subalgebra are realized in twisted sectors of non-free orbifold generators, turning the actions into free versions yields a reduction of the rank. Hence, we have a mechanism, which is an alternative to the CHL construction~\cite{CHL}, to decrease the rank of the gauge symmetry in heterotic string theory. 

As an example, we compactify toroidally the $\Z_2^5$-orbifold setup described in the previous section, and show that all gauge bosons generated by the fermions $\psi^i$ can be made massive by choosing free versions of all five orbifold generators, thus reducing the rank by 16 units. Then, we show in various ways how maximal rank can be recovered by keeping one of the generators not free. This will also be the occasion to stress subtleties in the manipulation of the characters of $SO(2n+1)$, especially the spinorial one.


\subsection{Reduction of the rank}
\label{redrank}

Let us consider the $\Spin$ heterotic string theory compactified on $T^5$, on which we implement a $\Z_2^5$ free-orbifold action.  The generators denoted $G_1^\f,\dots, G_5^\f$  act as $G_1,\dots, G_5$  on the fermions $\psi^i$, and as half-period shifts along the internal direction,
\be
G_{10-I}^\f = G_{10-I} \otimes \big( X^I\rightarrow X^I+\pi \big) \, ,~~\quad I\in\{5,\dots, 9\}~.
\label{g123f}
\ee
This $\Z_2^5$ action  was first introduced in Ref.~\cite{GKP}, though in a different base of generators, as well as in \Ref{so3so1} in presence of a spontaneous supersymmetry breaking. 
The one-loop partition function is given by 
\be
Z_5 = \frac{\bar V_8-\bar S_8}{(\sqrt{\Im \tau}\, \bar \eta\eta)^3}\, {1\over 2^5}\sum_{\substack{h_1,\dots,h_5 \\ g_1,\dots, g_5}} \frac{\Gamma_{5,5}\colvec[0.7]{h_1,\dots,h_5 \\ g_1,\dots,g_5}}{(\bar \eta\eta)^5}\,\frac{\Gamma^\Spin\colvec[0.7]{h_1,\dots,h_5 \\ g_1,\dots,g_5}}{\eta^{16}} ~ ,
\ee
where the lattice of zero modes of the compact coordinates $X^I$ can be considered either in  Lagrangian or Hamiltonian form~\cite{Kbook}. For arbitrary spacetime dimension $d$, it can be written as
\begin{align}
\Gamma_{10-d,10-d}&\colvec[0.7]{h_1,\dots,h_{10-d} \\ g_1,\dots,g_{10-d}}\nonumber \\
&\qquad ={\sqrt{\det G}\over (\Im \tau)^{10-d\over 2}}\sum_{\substack{n_{d},\dots,n_9 \\ \tilde m_{d},\dots, \tilde m_9}}e^{-{\pi\over \Im\tau}\left[\tilde m_I+g_{10-I}+(n_I+h_{10-I}) \tau](G+B)_{IJ}[\tilde m_J+g_{10-J}+(n_J+h_{10-J})\bar \tau\right]}\nonumber \\
&\qquad= \sum_{\substack{m_{d},\dots,m_9 \\ n_{d},\dots, n_9}}e^{-2i\pi g_{10-K}m_K}\;\bar q^{{1\over 4}P^\text{L}_IG^{-1}_{IJ}P^\text{L}_J}\; q^{{1\over 4}P^\text{R}_IG^{-1}_{IJ}P^\text{R}_J} ~ ,\esp
\label{ga}
 \end{align}
where we have defined 
\be
P^\text{L}_I=m_I+(B+G)_{IJ}\, (n_J+h_{10-J})~ , ~~\quad P^\text{R}_I=m_I+(B-G)_{IJ}\, (n_J+h_{10-J})~.
\label{lattice2}
\ee
In these formulas, $G_{IJ}$ and $B_{IJ}$ are the components of the internal metric and antisymmetric tensor, the momenta and winding numbers are denoted $m_I, n_I\in\Z$, and the sums over $\tilde m_I\in\Z$ are obtained by Poisson summation over the $m_I$'s.
 
From  \Eq{ga}, we see that any state such that some $m_I$ or $n_I+h_{10-I}$ is not vanishing has a mass squared that depends on the moduli $G_{IJ}, B_{IJ}$. They are therefore generically massive (as supersymmetry prevents the existence of tachyonic instabilities). Hence, in order to look for vector bosons massless for generic $G_{IJ}, B_{IJ}$, it is enough to focus on the untwisted sector at zero-momenta and zero-winding numbers,
\be
h_{10-I}=0~,\quad m_I=n_I=0~,~~\quad I\in\{5,\dots,9\}~.
\label{0-mass}
\ee
Restricting to these states in the partition function, we have in particular
\be
\begin{aligned}
 {1\over 2^5}\sum_{g_1,\dots ,g_5} \Gamma_{5,5}\colvec[0.7]{0,\dots,0 \\ g_1,\dots,g_5}\Big|_{\substack{m_5=\dots=m_9=0 \\ n_5=\dots= n_9=0}}\,\frac{\Gamma^\Spin\colvec[0.7]{0,\dots,0 \\ g_1,\dots,g_5}}{\eta^{16}} &= {1\over 2^5}\sum_{g_1,\dots ,g_5} \frac{\Gamma^\Spin\colvec[0.7]{0,\dots,0 \\ g_1,\dots,g_5}}{\eta^{16}}\\
 &=O_1^{32} + \text{massive}~ ,\esp
\end{aligned}
\label{diminu}
\ee
where we have used  \Eq{g1g2g3-3}. This shows that in the $\Z_2^5$ free-orbifold case, the gauge symmetry generated by the worldsheet fermions $\psi^i$ is trivial, $SO(1)^{32}$, with vanishing rank~\cite{so3so1}.

As seen in Ref.~\cite{ADLP}, the $SO(1)^{32}$ theory in five dimensions can also be realized as an orientifold model, which is dual to the heterotic picture~\cite{so3so1}.\footnote{In Ref.~\cite{ADLP, so3so1}, a spontaneous breaking of supersymmetry is also implemented by a  Scherk--Schwarz mechanism.\label{nonsysy}}${}^,$\footnote{See \Ref{ACP-1,ACP-2} for other orientifold models with reduced ranks but realizing the  $\N=2\to \N=0$ spontaneous breaking of supersymmetry in four dimensions.} One considers the type~I string compactified on $T^5$, and applies a T-duality on all periodic directions. The internal space becomes $\widetilde T^5/I_{56789}$, which is the dual torus of coordinates $\widetilde X^5,\dots, \widetilde X^9$ modded by the inversion $I_{56789}$,  $(\widetilde X^5,\dots, \widetilde X^9)\to -(\widetilde X^5,\dots, \widetilde X^9)$. There are 32 orientifold O5-planes, each of then located on one fixed point, as well as 32 D5-branes (dual to the D9-branes present before T-duality).  The $SO(1)^{32}$ model corresponds to distributing one D5-brane on each O5-plane. Because the configuration must be invariant under the inversion $I_{56789}$, all of these D5-branes have \emph{rigid} positions~\cite{Schwarz-rigid}. 

In the above open string picture, notice that sending to 0 the size of  one  direction, say  $\widetilde X^9$, 16 pairs of initially isolated D5-branes collapse, thus enhancing $\big(SO(1)\times SO(1)\big)^{16}\to SO(2)^{16}$.   Hence, decompactifying $X^9$ on the dual heterotic side should also yield an enhancement $\big(SO(1)\times SO(1)\big)^{16}\to SO(2)^{16}$. Maximal rank is therefore expected to be recovered at infinite distance in moduli space, which we can check. 
In fact, by proceeding  as in \Eq{diminu} with only one generator $G_1^\f$, one finds that the $\Spin$ theory compactified on $S^1$ and orbifolded by $G_1^\f$ realizes the $SO(32)\to SO(16)\times SO(16)$  breaking.  Moreover, it is not difficult to show that the $SO(32)$ gauge symmetry is recovered in the decompactification limit.\footnote{Physically, this means that the $SO(32)\to SO(16)\times SO(16)$ breaking is ``spontaneous''. The masses of the $SO(32)$ generators not in $SO(16)\times SO(16)$ are proportional to the inverse radius of $S^1$ and vanish in the decompactification limit.} It is then straightforward to compactify four directions $X^5,\dots,X^8$ and implement the actions of $G_2^\f, G_3^\f, G_4^\f, G_5^\f$, which  indeed break $SO(32)\to SO(2)^{16}$, as can be seen   by reasoning again as in \Eq{diminu}. 


\subsection{Restoration of the rank}
\label{orb1616tw}

As observed at the end of the previous subsection, a discontinuity of the rank may be encountered at infinite distance in moduli space. Alternatively, one may recover Cartan vector bosons by retrieving massless states in the twisted sector of non-free orbifold generators. 
In the following, we consider the  example of the $\Spin$ heterotic string theory compactified on the four-torus whose coordinates are $X^5,X^6,X^7,X^8$,  and orbifolded by the group $\Z_2^5$ generated by $G_1$ and the four free generators $G_2^\f, G_3^\f, G_4^\f ,G_5^\f$. As reviewed in Sect.~\ref{h8}, this model can also be seen as the $E_8\times E_8$ theory compactified on $T^4$ and modded by $\Z_2^4$ generated by $G_2^\f, G_3^\f, G_4^\f ,G_5^\f$. This setup turns out to reduce the initial $E_8\times E_8$ gauge symmetry to $U(1)^{16}$ in a non-trivial way. This will be shown by manipulating characters, and will be the opportunity to stress certain subtleties  associated with the spinorial character $S_{2n+1}$ of  $SO(2n+1)$.   

The partition function of the six-dimensional model is 
\be
Z_6 = \frac{\bar V_8-\bar S_8}{(\sqrt{\Im \tau}\, \bar \eta\eta)^4}\, {1\over 2^4}\sum_{\substack{h_2,\dots,h_5 \\ g_2,\dots, g_5}} \frac{\Gamma_{4,4}\colvec[0.7]{h_2,\dots,h_5 \\ g_2,\dots,g_5}}{(\bar \eta\eta)^4}\, {1\over 2}\sum_{h_1,g_1}\,\frac{\Gamma^\Spin\colvec[0.7]{h_1,\dots,h_5 \\ g_1,\dots,g_5}}{\eta^{16}} ~ ,
\ee
where the internal-torus lattice of zero modes is\footnote{Repeated indices $I,J,K$ are implicitly summed over $5,\dots,8$.} 
 \be
\Gamma_{4,4}\colvec[0.7]{h_2,\dots,h_5 \\ g_2,\dots,g_5}= \sum_{\substack{m_5,\dots,m_8 \\ n_5,\dots, n_8}}e^{-2i\pi g_{10-K}m_K}\;\bar q^{{1\over 4}P^\text{L}_IG^{-1}_{IJ}P^\text{L}_J}\; q^{{1\over 4}P^\text{R}_IG^{-1}_{IJ}P^\text{R}_J} ~ .
\label{ga2}
 \ee
As before, we are interested in the gauge symmetry encountered for generic values of the moduli $G_{IJ},B_{IJ}$. As explained above  \Eq{0-mass}, we may therefore concentrate our attention on the states satisfying 
\be
h_{10-I}=0~,~~m_I=n_I=0~,~~\quad I\in\{5,\dots,8\}~,
\ee
whose contributions to the partition function involve the sums
\be
\begin{aligned}
 {1\over 2^4}\sum_{g_2,\dots ,g_5} \Gamma_{4,4}\colvec[0.7]{0,\dots,0 \\ g_2,\dots,g_5}\Big|_{\substack{m_5=\dots=m_8=0 \\ n_5=\dots= n_8=0}}\,  {1\over 2}\sum_{h_1,g_1}&\frac{\Gamma^\Spin\colvec[0.7]{h_1,0,0,0,0 \\ g_1,g_2,g_3,g_4,g_5}}{\eta^{16}} \\
 &= {1\over 2^4}\sum_{g_2,\dots, g_5}  {1\over 2}\sum_{h_1,g_1} \frac{\Gamma^\Spin\colvec[0.7]{h_1,0,0,0,0 \\ g_1,g_2,g_3,g_4,g_5}}{\eta^{16}}~.
\end{aligned}
\ee
In the r.h.s., we have already shown in \Eq{g1g2g3-3} that the untwisted contribution $h_1=0$  yields a trivial subgroup $SO(1)^{32}$ of the full gauge symmetry generated by the fermions $\psi^i$. Hence, we are lead to analyze the gauge symmetry generated in the twisted sector $h_1=\half$. 


\paragraph{\em \boldmath Projection of the twisted sector $h_\alpha=\half \, \delta_{\alpha1}$:} 

To this end, we may consider  \Eq{twso32} which yields  
\be
\half\sum_{g_1}\frac{\Gamma^{\Spin}\colvec[0.7]{1/2 \\ g_1}}{\eta^{16}} =O_{16}S_{16}+S_{16}O_{16} ~ ,
\label{a}
\ee
and enforce successive projections onto $G_2\mbox{-}, \dots, G_5$-invariant root vectors.  
Decomposing all $SO(16)$ characters in terms of $SO(8)$ ones  by using \Eq{decomp1}, and applying the transformation rules~(\ref{Gtr}) for those associated with the fermions sensitive to the action of $G_2$, we obtain  
\be
\begin{aligned}
\frac{1}{2^2}\sum_{g_1,g_2}\frac{\Gamma^{\Spin}\colvec[0.7]{1/2,0 \\ g_1,g_2}}{\eta^{16}} &=\half \sum_{g_2}\Big[\big( O_8^2 + V_8(-1)^{2g_2}V_8\big) \big( S_8^2 + C_8(-1)^{2g_2}C_8\big)\\
&~~~~~~~~~\,+ \big( O_8^2 + V_8(-1)^{2g_2}V_8\big) \big( S_8^2 + C_8(-1)^{2g_2}C_8\big)\Big]\\
& =O_8^2S_8^2+S_8^2O_8^2+\text{massive}~,
\end{aligned}
\label{8}
\ee
where all terms involving $V_8V_8$ yield massive states only. Proceeding the same way to implement the projections onto the $G_3$- and $G_4$-invariant states, one obtains in a straightforward way 
\begin{align}
\frac{1}{2^4}&\sum_{g_1,\dots,g_4}\frac{\Gamma^{\Spin}\colvec[0.7]{1/2,0,0,0 \\ g_1,\dots,g_4}}{\eta^{16}}\nonumber \\
& =O_2^8\,\Big\{\big[S_2^8+S_2^4C_2^4 +S_2^2C_2^2S_2^2C_2^2+S_2^2C_2^4S_2^2 +(S^2_2,C^2_2)\to (S_2C_2,C_2S_2)\big]+\big[S_2\leftrightarrow C_2\big]\Big\}\nonumber \\
&~~~~\,+\Big\{\cdots\Big\}\, O_2^8+\text{massive} ~ ,\esp
\label{2}
\end{align}
where the content of each pair of braces is identical, with 16 terms.

In order to apply the projection onto the $G_5$-invariant states, we have to decompose the $SO(2)$ characters into $SO(1)$ ones. As already mentioned above the derivation of the untwisted sector contribution $h_1=0$ in \Eq{g1g2g3-3}, the necessary manipulations of $O_n$ and $V_n$ characters are irrespective of the parity of $n$. This is however not the case for the spinorial affine characters. Indeed, already at the level of the representations of $SO(n)$, there exist two irreducible spinorial representations of opposite chiralities when $n$ is even and a single one (non-chiral) when $n$ is odd. In practice, the characters $S_{2n}$ and $C_{2n}$ are equal since they differ only in the signs of vanishing contributions $\vartheta\colvec[0.7]{1/2\\1/2}^{n}$ in \Eq{defchar}. They may however be distinguished by considering their refined versions
\be
\begin{aligned}
S_{2n} (\nu_1,\dots,\nu_n|\tau)&= \half\left(\prod_{\alpha=1}^n {\vartheta\colvec[0.7]{ \overset{}1/2\\\overset{}0}(\nu_\alpha|\tau)\over \eta(\tau)} + \prod_{\alpha=1}^n {\vartheta\colvec[0.7]{1/2\\1/2}(\nu_\alpha|\tau)\over \eta(\tau)} \right)\, , \\
C_{2n} (\nu_1,\dots,\nu_n|\tau)&= \half\left(\prod_{\alpha=1}^n {\vartheta\colvec[0.7]{ \overset{}1/2\\ \overset{}0}(\nu_\alpha|\tau)\over \eta(\tau)} -\prod_{\alpha=1}^n  {\vartheta\colvec[0.7]{1/2\\1/2}(\nu_\alpha|\tau)\over \eta(\tau)} \right)\, , 
\end{aligned}
\ee
where $(\nu_1,\dots,\nu_n)$ parametrizes the Cartan subalgebra~\cite{Kbook}. In the case of $SO(2n+1)$ characters, there is no such distinction since there is always one factor $\vartheta\colvec[0.7]{1/2\\1/2}^{\half}(0|\tau)=0$ that cannot be ``refined'', so that
\be
S_{2n+1} (\nu_1,\dots,\nu_n|\tau)= {1\over \sqrt{2}}\left(\prod_{\alpha=1}^n {\vartheta\colvec[0.7]{ \overset{}1/2\\ \overset{}0}(\nu_\alpha|\tau)\over \eta(\tau)} \right)\sqrt{{\vartheta\colvec[0.7]{ \overset{}1/2\\ \overset{}0}(0|\tau)\over \eta(\tau)}}~ .  
\ee
In particular, we may write $S_2=S_1^2$ and $C_2=S_1^2$ by applying identities given in \Eq{decomp1}. Notice however that in \Eq{decomp1} we have kept track of all contributions involving $\vartheta\colvec[0.7]{1/2\\1/2}$'s by introducing a vanishing quantity $\Delta_{2n+1}$ defined in \Eq{delta}. This can be useful if one wants to reconstruct without ambiguity the partition function in terms of $SO(2n)$ characters rather than $SO(2n+1)$. But more importantly, we will see at the end of the present section that such terms are useful in some derivations.  In any case, whether we keep these extra terms or not, we obtain  from \Eq{Gtr}  that flipping the boundary conditions along the cycle $[0,\tau]$ of all fermions sensitive to the action of $G_5$ transforms 
\be
\begin{aligned}
S_2=S_1^2+\Delta_1^2~~\longrightarrow~~ S_1\,  i^\half\Delta_1+\Delta_1\,  i^\half S_1 =0~,\\
C_2=S_1^2-\Delta_1^2~~\longrightarrow~~ S_1\,  i^\half\Delta_1-\Delta_1\,  i^\half S_1 =0~.
\end{aligned}
\label{81}
\ee
As a result, all non-trivial contributions arise for $g_5=0$, so that
\be
\begin{aligned}
\frac{1}{2^5}&\sum_{g_1,\dots,g_5}\frac{\Gamma^{\Spin}\colvec[0.7]{1/2,0,\dots,0 \\ g_1,\dots,g_5}}{\eta^{16}}\\
&= \half \left[\big(O_1^2+V_1^2)^8\,\Big\{S_1^{16}+\mbox{15 similar terms}\Big\}+\Big\{\cdots\Big\}\, \big(O_1^2+V_1^2)^8\right]+\text{massive}\\
&= 8\, O_1^{16}S_1^{16}+8\, S_1^{16}O_1^{16}+\text{massive}~.\esp
\end{aligned}
\label{res}
\ee
From the $q$-expansions given in \Eq{decomp1}, we see that each of the above 16 terms contribute a massless state or, rather, a massless vector multiplet in the full theory. 
However, \Eq{res} is telling us that these vector multiplets are spinors of  the trivial $SO(1)^{32}$ subgroup generated in the untwisted sector, which is an empty statement. In particular, it is not clear at this stage whether the 16 vector multiplets are Abelian or non-Abelian. The goal of the next paragraph is to answer this question. 


\paragraph{\em \boldmath Projection of the twisted Fock space $h_\alpha=\half \, \delta_{\alpha1}$:} 

In order to figure out the self interactions of the 16 gauge bosons generated in the twisted sector $h_1=\half, h_2=\dots=h_5=0$, let us take a look at the associated Fock space. To this end, it is convenient to define  a complex basis of fermions
\be
\Psi^u={\psi^{2u-1}+i\psi^{2u}\over \sqrt{2}}~,\quad \Psi^{u\dagger}={\psi^{2u-1}-i\psi^{2u}\over \sqrt{2}}~,~~\quad u\in\{1,\dots,16\}~.
\ee
In the right-moving Ramond $h_\alpha=\half \,\delta_{\alpha1}$ sector, which is associated with the characters $S_{16}O_{16}$ in \Eq{a}, $\Psi^1,\dots,\Psi^{8}$ have integer modes, while $\Psi^9,\dots,\Psi^{16}$ have half-integer modes. Similarly, in the Neveu--Schwarz $h_\alpha=\half \,\delta_{\alpha1}$ sector, which corresponds to $O_{16}S_{16}$, the fermions $\Psi^9,\dots,\Psi^{16}$ have integer modes, while $\Psi^1,\dots,\Psi^{8}$ have half-integer modes. Let us focus on the Ramond sector, keeping in mind that the Neveu--Schwarz sector can be analyzed in the same  way. In the Ramond sector, the algebra of the zero modes 
\be
\{\Psi^u_0,\Psi^{v\dagger}_0\}=\delta^{uv}~,\quad \{\Psi^u_0,\Psi^{v}_0\}=\{\Psi^{u\dagger}_0,\Psi^{v\dagger}_0\}=0~,~~\quad u\in\{1,\dots,8\}~, 
\label{alg}
\ee
implies the vacuum to be a spinor with $2^8$ components. One of them, denoted $\left|\mbox{$+{1\over 2}$},\dots,\mbox{$+{1\over 2}$}\right\rangle$, vanishes under the action of the $\Psi^{u\dagger}_0$'s, and generates all other components by applying the  $\Psi^{u}_0$'s,
\be
\begin{aligned}
&\left|\mbox{${s_1\over 2}$},\dots,\mbox{${s_8\over 2}$}\right\rangle=(\Psi_{0}^1)^{1-s_1\over 2}\cdots (\Psi_{0}^8)^{1-s_8\over 2}\left|\mbox{$+{1\over 2}$},\mbox{$+{1\over 2}$},\mbox{$+{1\over 2}$},\mbox{$+{1\over 2}$},\mbox{$+{1\over 2}$},\mbox{$+{1\over 2}$},\mbox{$+{1\over 2}$},\mbox{$+{1\over 2}$}\right\rangle\,, ~~ s_1,\dots,s_8=\pm1~.
\end{aligned}
\label{deef}
\ee
Imposing $\left|\mbox{$+{1\over 2}$},\dots,\mbox{$+{1\over 2}$}\right\rangle$ to be normalized, the relations~(\ref{alg}) can be used to show that the $\left|\mbox{${s_1\over 2}$},\dots,\mbox{${s_8\over 2}$}\right\rangle$ form an orthonormal basis.  
Restricting to the spinorial representation of $SO(16)$ that corresponds to the low lying states of the characters $S_{16}O_{16}$ amounts to imposing $\prod_{u=1}^8 s_u=+1$, which selects $2^8/2$ choices of signs.\footnote{For $SO(2n)$, the spinorial and anti-spinorial representations, which are associated with the characters $S_{2n}$ and $C_{2n}$, are obtained by imposing $\prod_{u=1}^n s_u=+1$ and $\prod_{u=1}^n s_u=-1$, respectively. } 

Denoting for convenience $\Psi^{u\dagger}_0\equiv \Psi^{u,+1}_0$ and 
$\Psi^u_0\equiv\Psi^{u,-1}_0$, the components of the vacuum satisfy
\be
\left|\mbox{${s_1\over 2}$},\dots,\mbox{${s_8\over 2}$}\right\rangle=\Big(\prod_{u=1}^8 \Psi_0^{u,s_u}\, \Psi_0^{u,-s_u}\Big)\left|\mbox{${s_1\over 2}$},\dots,\mbox{${s_8\over 2}$}\right\rangle\,.
\label{truc}
\ee
Under any symmetry $G_\alpha$, where $\alpha=2,3$ or 4, they must be mapped into low lying states, \ie linear combinations among themselves. Hence, we have 
\be
\begin{aligned}
G_\alpha\left|\mbox{${s_1\over 2}$},\dots,\mbox{${s_8\over 2}$}\right\rangle&= \sum_{t_1,\dots,t_8}c_{t_1,\dots,t_8}^{(\alpha)}\left|\mbox{${t_1\over 2}$},\dots,\mbox{${t_8\over 2}$}\right\rangle\\
&=G_\alpha \bigg[\Big(\prod_{u=1}^8 \Psi_0^{u,s_u}\, \Psi_0^{u,-s_u}\Big)\left|\mbox{${s_1\over 2}$},\dots,\mbox{${s_8\over 2}$}\right\rangle\bigg]\\
&=\sum_{t_1,\dots,t_8}c_{t_1,\dots,t_8}^{(\alpha)}\Big(\prod_{u=1}^8 \Psi_0^{u,s_u}\, \Psi_0^{u,-s_u}\Big)\left|\mbox{${t_1\over 2}$},\dots,\mbox{${t_8\over 2}$}\right\rangle=c_{s_1,\dots,s_8}^{(\alpha)} \left|\mbox{${s_1\over 2}$},\dots,\mbox{${s_8\over 2}$}\right\rangle,
\end{aligned}
\ee
where we have used in the second line the fact that all $\Psi_0^{u,s_u}\, \Psi_0^{u,-s_u}$ are invariant under $G_\alpha$. Because the square of  $G_\alpha$ is the identity, we have  $(c_{s_1,\dots,s_8}^{(\alpha)})^2=1$, which shows that all components $\left|\mbox{${s_1\over 2}$},\dots,\mbox{${s_8\over 2}$}\right\rangle$ diagonalize $G_2,G_3$ and $G_4$.  

It turns out that $\left|\mbox{$+{1\over 2}$},\dots,\mbox{$+{1\over 2}$}\right\rangle$ has eigenvalues $c_{+1,\dots,+1}^{(\alpha)}=+1$ \ie is invariant under  these generators. For $G_2$, this is shown in \Eq{8}, as it is one of the low lying states of the monomial $S_8^2O_8^2$. Similarly, $\left|\mbox{$+{1\over 2}$},\dots,\mbox{$+{1\over 2}$}\right\rangle$  is a low lying state of the characters $S_4^4O_4^4$ and $S_2^8O_2^8$, which survive the projections associated with $G_3$ and $G_4$ (see \Eq{2}). It is then straightforward to identify from the definition~(\ref{deef}) all other low lying states of $S_{16}O_{16}$ that are invariant under $G_2,G_3,G_4$,  
\be
\label{rootssu2}
\begin{aligned}
&\left|\mbox{$+{1\over 2}$},\mbox{$+{1\over 2}$},\mbox{$+{1\over 2}$},\mbox{$+{1\over 2}$},\mbox{$+{1\over 2}$},\mbox{$+{1\over 2}$},\mbox{$+{1\over 2}$},\mbox{$+{1\over 2}$}\right\rangle\,,~~\quad\left|\mbox{$+{1\over 2}$},\mbox{$+{1\over 2}$},\mbox{$+{1\over 2}$},\mbox{$+{1\over 2}$},\mbox{$-{1\over 2}$},\mbox{$-{1\over 2}$},\mbox{$-{1\over 2}$},\mbox{$-{1\over 2}$}\right\rangle\,,\espD
 \\
&\left|\mbox{$+{1\over 2}$},\mbox{$+{1\over 2}$},\mbox{$-{1\over 2}$},\mbox{$-{1\over 2}$},\mbox{$+{1\over 2}$},\mbox{$+{1\over 2}$},\mbox{$-{1\over 2}$},\mbox{$-{1\over 2}$}\right\rangle\,,~~\quad\left|\mbox{$+{1\over 2}$},\mbox{$+{1\over 2}$},\mbox{$-{1\over 2}$},\mbox{$-{1\over 2}$},\mbox{$-{1\over 2}$},\mbox{$-{1\over 2}$},\mbox{$+{1\over 2}$},\mbox{$+{1\over 2}$}\right\rangle\,,\espD
 \\
 &\left|\mbox{$+{1\over 2}$},\mbox{$-{1\over 2}$},\mbox{$+{1\over 2}$},\mbox{$-{1\over 2}$},\mbox{$+{1\over 2}$},\mbox{$-{1\over 2}$},\mbox{$+{1\over 2}$},\mbox{$-{1\over 2}$}\right\rangle\,,~~\quad \left|\mbox{$+{1\over 2}$},\mbox{$-{1\over 2}$},\mbox{$+{1\over 2}$},\mbox{$-{1\over 2}$},\mbox{$-{1\over 2}$},\mbox{$+{1\over 2}$},\mbox{$-{1\over 2}$},\mbox{$+{1\over 2}$}\right\rangle\,,\espD
 \\
 &\left|\mbox{$+{1\over 2}$},\mbox{$-{1\over 2}$},\mbox{$-{1\over 2}$},\mbox{$+{1\over 2}$},\mbox{$+{1\over 2}$},\mbox{$-{1\over 2}$},\mbox{$-{1\over 2}$},\mbox{$+{1\over 2}$}\right\rangle\,,~~\quad\left|\mbox{$+{1\over 2}$},\mbox{$-{1\over 2}$},\mbox{$-{1\over 2}$},\mbox{$+{1\over 2}$},\mbox{$-{1\over 2}$},\mbox{$+{1\over 2}$},\mbox{$+{1\over 2}$},\mbox{$-{1\over 2}$}\right\rangle\,,
 \\
 &\mbox{and all components obtained by changing $+{1\over 2}\leftrightarrow-{1\over 2}$}~.\esp
\end{aligned}
\ee
Consistently, they match the 16 monomials $\{S_2^8+\cdots\}\,O_2^8$ of \Eq{2} under the dictionary $(S_2,C_2)\to (\mbox{$+{1\over 2}$},\mbox{$-{1\over 2}$})$. After implementation of the projections onto $G_2$-, $G_3$-, $G_4$-states, the untwisted sector contributes a subgroup $SO(2)^{16}$ of the gauge symmetry (see \Eq{g1g2g3-2}). Moreover, the states in \Eq{rootssu2} are labelled by their weights \ie charges under the first $SO(2)^8$  factor of $SO(2)^{16}$. The key point is that the weight vectors of the 8 states explicitly listed in \Eq{rootssu2} are all orthogonal to each other. Hence, the associated Dynkin diagram is composed of 8 disconnected dots, which corresponds to $SU(2)^8$. 

We are left with one task, which is to implement the projection onto the $G_5$-invariant states. The generator $G_5$ being a symmetry, it maps the low lying states into linear combinations among themselves. Hence, 
\begin{align}
G_5\left|\mbox{${s_1\over 2}$},\dots,\mbox{${s_8\over 2}$}\right\rangle&= \sum_{t_1,\dots,t_8}c_{t_1,\dots,t_8}^{(5)}\left|\mbox{${t_1\over 2}$},\dots,\mbox{${t_8\over 2}$}\right\rangle\nonumber \\
&=G_5 \bigg[\Big(\prod_{u=1}^8 \Psi_0^{u,s_u}\, \Psi_0^{u,-s_u}\Big)\left|\mbox{${s_1\over 2}$},\dots,\mbox{${s_8\over 2}$}\right\rangle\bigg]\label{3}\\
&=\sum_{t_1,\dots,t_8}c_{t_1,\dots,t_8}^{(5)}\Big(\prod_{u=1}^8 \Psi_0^{u,-s_u}\, \Psi_0^{u,s_u}\Big)\left|\mbox{${t_1\over 2}$},\dots,\mbox{${t_8\over 2}$}\right\rangle=c_{-s_1,\dots,-s_8}^{(5)} \left|\mbox{$-{s_1\over 2}$},\dots,\mbox{$-{s_8\over 2}$}\right\rangle\, ,\nonumber 
\end{align}
where we have used the fact that $G_5$ exchanges all $\Psi_0^{u\dagger}\leftrightarrow \Psi^{u}_0$. Because $G_5^2$ is the identity, we obtain that $c^{(5)}_{-s_1,\dots,-s_8}c^{(5)}_{s_1,\dots,s_8}=1$ (no sum over the indices). Moreover, $G_5$ being Hermitian, taking  in \Eq{3} the Hermitian product with $\left\langle\mbox{$-{s_1\over 2}$},\dots,\mbox{$-{s_8\over 2}$}\right|$  yields $(c^{(5)}_{s_1,\dots,s_8})^*=c^{(5)}_{-s_1,\dots,-s_8}$. All coefficients are therefore pure phases, $c^{(5)}_{s_1,\dots,s_8}=e^{2i\theta_{s_1,\dots,s_8}}$, where $\theta_{-s_1,\dots,-s_8}=-\theta_{s_1,\dots,s_8}$. As a result, the states 
\be
{1\over \sqrt{2}}\,\Big(e^{i\theta_{s_1,\dots,s_8}}\left|\mbox{${s_1\over 2}$},\dots,\mbox{${s_8\over 2}$}\right\rangle\pm e^{i\theta_{-s_1,\dots,-s_8}}\left|\mbox{$-{s_1\over 2}$},\dots,\mbox{$-{s_8\over 2}$}\right\rangle\!\Big)
\ee
form an orthonormal base that diagonalizes $G_5$ and among which the ``+'' combinations are invariant. In fact, we may have worked from the beginning with the vacuum states  $e^{i\theta_{s_1,\dots,s_8}}\left|\mbox{${s_1\over 2}$},\dots,\mbox{${s_8\over 2}$}\right\rangle$ rather than $\left|\mbox{${s_1\over 2}$},\dots,\mbox{${s_8\over 2}$}\right\rangle$, which amounts to absorb the phases in the definitions of the ``bra''. In that case, the last projection onto $G_5$-invariant states has two effects. Firstly, it eliminates from the untwisted sector (the right-moving Neveu--Schwarz $h_1=\dots=h_5=0$ sector) the Cartan generators of $SU(2)^8$. Secondly, it identifies all positive roots of $SU(2)^8$, which correspond to the ``bra'' explicitly displayed in \Eq{rootssu2},  with their opposite negative roots, thus yielding a system of 8 Abelian $U(1)$ generators. 

Taking into account both Ramond and Neveu--Schwarz $h_\alpha=\half \, \delta_{\alpha1}$ sectors, the full gauge symmetry generated by the fermions $\psi^i$  in the $\Spin$  theory compactified on $T^4$ parametrized by $X^5,X^6,X^7,X^8$  and orbifolded by the group generated by $G_1$ and $G_2^\f, G_3^\f, G_4^\f ,G_5^\f$ is $U(1)^{16}$, which is of maximal rank. 


\paragraph{\em \boldmath Twisting the invariant states of the sector $h_\alpha=0$:} 

To derive \Eq{res}, we have considered the twisted sector $h_\alpha=\half\, \delta_{\alpha1}$ of the lattice, $\Gamma^{\Spin}\colvec[0.7]{1/2 \\ 0}$, on which we have applied projections onto $G_2\mbox{-}, \dots, G_5$-invariant weights. In the following, we recover the result by reversing the operations. We start with the untwisted sector $h_\alpha=0$ \ie the lattice $\Gamma^{\Spin}$, implement the five projections, and only then change the boundary conditions along the cycle $[0,1]$ in order to  switch to the twisted sector $h_\alpha=\half \, \delta_{\alpha1}$. This turns out to be an interesting exercise for the following reason:

Under a change of boundary conditions along the cycle $[0,1]$, vanishing terms involving factors $\vartheta\colvec[0.7]{1/2 \\ 1/2}^\half$ in a partition function are related to non-trivial contributions involving $\vartheta\colvec[0.7]{0 \\ 1/2}^\half$. When a partition function is expressed in terms of characters $S_{2n}$ and/or $C_{2n}$, this mapping between untwisted and twisted sectors is always satisfied, since all Jacobi modular forms $\vartheta\colvec[0.7]{1/2 \\ 1/2}^\half$ are taken into account to provide physical information on the chirality of the spectrum. However, when the partition function is expressed in terms of $SO(2n+1)$ spinorial characters $S_{2n+1}$, there is no justification from the point of view of representation theory to keep track of the contributions involving $\vartheta\colvec[0.7]{1/2 \\ 1/2}^\half$. As illustrated in the following, we want to stress that it is nevertheless mandatory to consider such terms when the mapping between untwisted and twisted sectors is to be used.\footnote{Notice that the link between $\vartheta\colvec[0.7]{0 \\ 1/2}$ and $\vartheta\colvec[0.7]{1/2 \\ 1/2}$ is to be distinguished with modular invariance, since terms involving $\vartheta\colvec[0.7]{1/2 \\ 1/2}$ are mapped into themselves under $SL(2,\Z)$ transformations. Indeed, vanishing terms are (obviously) modular invariant.}

Starting with $\Gamma^\Spin=O_{32}+S_{32}$, we can use iteratively the decompositions formula~(\ref{decomp1}) and transformation rules~(\ref{Gtr}) to implement the projections associated with the generators $G_1,G_2,G_3$. This yields
\begin{align}
\frac{1}{2^3}\sum_{g_1,g_2,g_3}\frac{\Gamma^{\Spin}\colvec[0.7]{0,0,0 \\ g_1,g_2,g_3}}{\eta^{16}} &=\Big\{\big[O_4^8+O_4^4V_4^4+O_4^2V_4^2O_4^2V_4^2+O_4^2V_4^4O_2^2+O_4^2\leftrightarrow V_4^2\big]\label{5}\\
&~~~~~+\big[(O_4^2,V_4^2)\to (V_4O_4,O_4V_4)\big]\Big\}+\Big\{(O_4,V_4)\to (C_4,S_4)\Big\}\;,\nonumber 
\end{align}
where there are 16 monomials in each pair of braces. Notice that contrary to \Eq{g1g2g3}, we write explicitly all characters, including those that yield only massive contributions, as we are ultimately interested in mapping them into $h_\alpha=\half\, \delta_{\alpha1}$-twisted sector contributions. Let us concentrate first on the 16 monomials $X^{(1)}_4\cdots X^{(8)}_4$ where all  factors are either $O_4$ or $V_4$. As seen in the sequel,  the last four factors $X^{(5)}_4\cdots X^{(8)}_4$ generate a multitude of terms $S_1^{16}$ after implementation of the last two projections and the twist by $G_1$. To obtain massless states, the four first factors $X^{(1)}_4\cdots X^{(4)}_4$ must therefore contribute prefactors $O_1^{16}$, which do note contain characters $V_1$. This constraint imposes $X^{(1)}_4\cdots X^{(4)}_4=O_4^4$, which is satisfied only by the first two monomials in the r.h.s. of \Eq{5}.   
Implementing the projection associated with $G_4$ on the terms $O_4^4(O_4^4+V_4^4)$,  one obtains
\begin{align}
&\left.\frac{1}{2^4}\sum_{g_1,\dots,g_4}\frac{\Gamma^{\Spin}\colvec[0.7]{0,\dots ,0\\ g_1,\dots,g_4}}{\eta^{16}}\right|_{\text{order-8 monomials in $O_4,V_4$}} \\
&~~~~~~~~~=O_2^8\Big(O_2^8+\mbox{15 other order-8 monomials in $O_2,V_2$}\Big) +\mbox{massive after twisting by $G_1$}~,\esp\nonumber 
\end{align}
while the last projection onto $G_5$-invariant weights yields
\be\begin{aligned}
&\left.\frac{1}{2^5}\sum_{g_1,\dots,g_5}\frac{\Gamma^{\Spin}\colvec[0.7]{0,\dots ,0\\ g_1,\dots,g_5}}{\eta^{16}}\right|_{\text{order-8 monomials in $O_4,V_4$}} \\
&\qquad\qquad\qquad  ~~~~=O_1^{16}\Big(O_1^{16}+\mbox{$\big({2^8\over 2}\times 16-1\big)$ other order-16 monomials in $O_1,V_1$}\Big)\esp \\
&~~~~\qquad\qquad\qquad\quad \,+ \mbox{massive after twisting by $G_1$}~. \esp
\end{aligned}
\label{6}
\ee
We are now ready to flip the boundary conditions along the cycle $[0,1]$ of the fermions $\psi^i$, $i\in\{17,\dots, 32\}$. The overall factor $O_1^{16}$ in \Eq{6} is invariant while all $O_1$ and $V_1$ characters in the parenthesis are transformed according to \Eq{Gtwist}, which yields
\be
O_1~~\longrightarrow~~ {S_1+i\, \Delta_1\over \sqrt{2}}={S_1\over \sqrt{2}}~,\qquad V_1~~\longrightarrow~~ {S_1-i\, \Delta_1\over \sqrt{2}}={S_1\over \sqrt{2}}~.
\ee
Notice that it is safe to omit all vanishing $\Delta_1$ terms, which do not contain information on the twisted spectrum. 
As a result, we find
\begin{align}
\left.\frac{1}{2^5}\sum_{g_1,\dots,g_5}\frac{\Gamma^{\Spin}\colvec[0.7]{1/2,0,\dots ,0\\ g_1,\dots,g_5}}{\eta^{16}}\right|_{\text{order-8 monomials in $O_4,V_4$}} & =O_1^{16}\,\Big({S_1\over\sqrt{2}}\Big)^{16}\times {2^8\over 2}\times 16 + \mbox{massive}\nonumber  \\
&=8\, O_1^{16}S_1^{16}+\text{massive}~,
\end{align}
which reproduces the first monomial in \Eq{res}.

Let us move on the 16 monomials in \Eq{5} that contain characters $C_4$ or $S_4$. Implementing the projection associated with $G_4$, we obtain 
\begin{align}
&\left.\frac{1}{2^4}\sum_{g_1,\dots,g_4}\frac{\Gamma^{\Spin}\colvec[0.7]{0,\dots ,0\\ g_1,\dots,g_4}}{\eta^{16}}\right|_{\text{order-8 monomials in $C_4,S_4$}}\label{10} \\
&~~~~~~~~~~~~~~~=(S_2C_2)^8+\mbox{$\big({2^8\over 2}\times 16-1\big)$ other order-16 monomials in $S_2,C_2$} ~.\esp\nonumber 
\end{align}
As seen in \Eq{81}, flipping the boundary conditions along the cycle $[0,\tau]$ of all fermions sensitive to the action of $G_5$ transforms the characters  $S_2,C_2$ into $S_1\,  i^\half\Delta_1\pm \Delta_1\,  i^\half S_1$. Moreover, when this is done on the 8 first characters $S_2$ or $C_2$ of each monomial in \Eq{10}, these vanishing factors are invariant under the action of $G_1$ that flips boundary conditions along $[0,1]$. Hence, we are free to omit all contributions $g_5=1$ when implementing the projection onto $G_5$-invariant states, which yields
 \begin{align}
&\left.\frac{1}{2^5}\sum_{g_1,\dots,g_5}\frac{\Gamma^{\Spin}\colvec[0.7]{0,\dots ,0\\ g_1,\dots,g_5}}{\eta^{16}}\right|_{\text{order-8 monomials in $C_4,S_4$}}\label{11}\nonumber  \\
&~~~~~~~~~~~~~~~=\half\,\Big\{\big[(S_1^2+\Delta_1^2)(S_1^2-\Delta_1^2)\big]^8\\
&~~~~~~~~~~~~~~~~~~~~~~~+\mbox{$\big({2^8\over 2}\times 16-1\big)$ other order-16 monomials in $S_1^2+\Delta_1^2,S_1^2-\Delta_1^2$} \Big\}\,.\esp\nonumber 
\end{align}
As an example, let us focus on the first of the  ${2^8\over 2}\times 16$ terms. Flipping the boundary conditions along the cycle $[0,1]$ of the fermions sensitive to the action of $G_1$, this term transforms according to \Eq{Gtwist}, 
\be
\begin{aligned}
\half \big[(S_1^2+\Delta_1^2)(S_1^2-\Delta_1^2)\big]^8&~~\longrightarrow~~\half \big[(S_1^2+\Delta_1^2)(S_1^2-\Delta_1^2)\big]^4\\
&\hspace{-3cm}\times\bigg\{\Big[\Big({O_1+V_1\over \sqrt{2}}\Big)^2+\Big(i^{-\half}{O_1-V_1\over \sqrt{2}}\Big)^2\Big]\Big[\Big({O_1+V_1\over \sqrt{2}}\Big)^2+\Big(i^{-\half}{O_1-V_1\over \sqrt{2}}\Big)^2\Big]\bigg\}^4\\
&~~\longrightarrow~~{2^4\over 2} \, S_1^{16}\Big({O_1\over \sqrt{2}}\Big)^{16}+\mbox{massive}~.
\end{aligned}
\label{12}
\ee
Notice that we are free to  set $\Delta_1=0$ in the first 8 factors $(S_1^2\pm \Delta_1^2)$, while it is mandatory to keep track of all $\Delta_1$'s in the last 8 factors $(S_1^2\pm \Delta_1^2)$ because they transform into non-trivial contributions of the twisted sector $h_\alpha=\half\, \delta_{\alpha1}$. In general, the monomials in \Eq{10} containing $n$ characters $S_2$ and $8-n$ characters $C_2$ in their last 8 factors contribute as shown in \Eq{12} up to a multiplicative coefficient $i^n$.  
Due to a symmetry that exchanges $S_2\leftrightarrow C_2$, the imaginary terms ($n$ odd) cancel each other and we are left with contributions for $n=0,2,4,6,8$. By counting how many terms can be generated for each $n$ from the 16 monomials involving $C_4,S_4$ in \Eq{5}, we find that 
\be
\begin{aligned}
&\left.\frac{1}{2^5}\sum_{g_1,\dots,g_5}\frac{\Gamma^{\Spin}\colvec[0.7]{1/2,0,\dots ,0\\ g_1,\dots,g_5}}{\eta^{16}}\right|_{\text{order-8 monomials in $C_4,S_4$}} \\
&\qquad \qquad\qquad \qquad ={2^4\over 2} \, S_1^{16}\Big({O_1\over \sqrt{2}}\Big)^{16} \times {2^4\over 2}\,\Big(2-56+140-56+2\Big)+\mbox{massive}  \\
&\qquad \qquad \qquad \qquad=8\, S_1^{16}O_1^{16}+\text{massive}~,\esp
\end{aligned}
\ee
which reproduces the second monomial in \Eq{res}. Had we omitted the $\Delta_1$'s responsible for the contributions  $i^{-\half}(O_1-V_1)/\sqrt{2}$ once twisted by $G_1$, we would have found half of the correct answer, $4\, S_1^{16}O_1^{16}$. 


\section{From Wilson line backgrounds to orbifolds}
\label{wl}

In this section, we would like to stress some links between Wilson line backgrounds and orbifold actions. In quantum field theory, switching on moduli in the Coulomb branch of a gauge theory implies a reduction of the dimension of the gauge group but the preservation of its rank. In the context of string theory,  Wilson-line backgrounds are realized by switching one marginal deformations of the worldsheet conformal field theory. However, it turns out that Wilson line deformations around backgrounds  can alternatively be described by implementing orbifold actions, which are free in order not to generate massless gauge bosons associated with extra generators in twisted sectors.\footnote{These extra generators can be of non-Cartan type, as seen in Sect.~\ref{h8} \ref{8h}. They can also be of Cartan type, as seen in various ways in Sect.~\ref{orb1616tw}. } In the following, we first illustrate this fact for Wilson-line backgrounds corresponding to $\Z_2$ free orbifold actions in presence of discrete torsion~\cite{torsion-1,torsion-2}.  Then, we show that the correspondence of the two approaches can be extended to cases where the free orbifold actions reduce the rank, as seen for instance in Sect~\ref{redrank}. When this is so, however,  the deformations are discrete and  no longer vacuum expectation values of moduli fields, as they actually transform the initial model into another model belonging to a distinct component of the moduli space. 


\subsection{\boldmath Wilson lines of the ${\rm Spin}(32)/\pmb{\Z}_2$  theory on $T^{10-d}$} 
\label{wlT}

The master formula we are going to use extensively is the expression of the Narain lattice~\cite{Narain1,Narain2} of the $\Spin$ heterotic string theory compactified toroidally down to $d$~dimensions~\cite{421,CP}. The latter is a moduli-dependent, even, self-dual and Lorentzian lattice of signature $(10-d,26-d)$, which may be written as
\be
\begin{aligned}
\Gamma_{10-d,26-d}&=\half\sum_{\gamma,\delta}\Gamma_{10-d,26-d}\colvec[0.7]{\gamma \\ \delta}\;,~~\quad \where\\
\Gamma_{10-d,26-d}\colvec[0.7]{\gamma \\ \delta}(G,B,\vec Y)&={\sqrt{\det G}\over (\Im \tau)^{10-d\over 2}}\sum_{\substack{n_d,\dots,n_9 \\ \tilde m_d,\dots, \tilde m_9}} e^{-{\pi\over \Im \tau}(\tilde m_I+n_I\tau)(G+B)_{IJ}(\tilde m_J+n_J\bar \tau)} \\
&\hspace{3cm} \times e^{i\pi n_I\vec Y_{I} \cdot \left(2\delta \vec \II_{16}-\tilde m_J\vec Y_{J}\right)} \, \prod_{u=1}^{16} \vartheta\colvec[0.7]{\gamma -n_KY_{Ku}\\ \delta-\tilde m_LY_{Lu}}~ .
\end{aligned}
\label{block}
\ee
In this expression, $\vec Y_I$ is a vector with real entries $Y_{Iu}$, $u\in\{1,\dots 16\}$, and $\vec \II_{16}$ is the 16-vector whose components are all equal to 1. We choose to write the contribution associated with the zero modes of the internal torus $T^{10-d}$ in Lagrangian form. The moduli space of marginal deformations of the Narain lattice has real dimension $(10-d)\times(26-d)$. Hence, it is fully parametrized by the internal metric and antisymmetric tensor, $(G+B)_{IJ}$, $I,J\in\{d,\dots, 9\}$, together with $Y_{Iu}$, $u\in\{1,\dots,16\}$. To show that modular invariance of any partition function involving $\Gamma_{10-d,26-d}\colvec[0.7]{\gamma \\ \delta}$ is not spoiled by these deformations, one can check  that the transformation properties under the $SL(2,\Z)$ generators are independent of $(G+B)_{IJ}$ and $Y_{Ku}$, 
\be
\begin{aligned}
\tau\to \tau+1\quad &\Longrightarrow \quad (\gamma,\delta)\to \Big(\gamma,\delta+\gamma-\half\Big)\;,\\
\tau\to-{1\over  \tau}\quad &\Longrightarrow \quad (\gamma,\delta)\to (\delta,-\gamma)~.
\end{aligned}
\ee
However, because the above transformations send $\gamma,\delta$ in $\Z\cup(\Z+\half)$ rather than $\{0,\half\}$, one also has to check that changing $\gamma\to \gamma+1$ or $\delta\to \delta+1$ affects $\Gamma_{10-d,26-d}\colvec[0.7]{\gamma \\ \delta}$ in a way that is  independent of $(G+B)_{IJ}$ and $Y_{Ku}$. This turns out to be trivially the case since 
\be
\Gamma_{10-d,26-d}\colvec[0.7]{\gamma+1 \\ \delta}(G,B,Y)=\Gamma_{10-d,26-d}\colvec[0.7]{\gamma \\ \delta+1}(G,B,Y)=\Gamma_{10-d,26-d}\colvec[0.7]{\gamma \\ \delta}(G,B,Y)~.
\ee
Because the quantum numbers of the degrees of freedom described by the partition function are discrete, they are independent of the continuous deformations. On the contrary, nothing protects the associated masses to depend on the moduli.\footnote{Technically, invariance of the partition function under $\tau\to \tau+1$ can be used to show that the dependence on $(G+B)_{IJ}$ and $Y_{Iu}$ disappears completely from the level-matching condition $\bar L_0-\half=L_0-1$, where $\bar L_0, L_0$ are the zero modes of the left- and right-moving Virasoro generators. On the contrary, each side of this equality is proportional to the mass squared operator and depends on the moduli.}  In a maximally supersymmetric heterotic string model (\eg $\N=4$ in four dimensions), varying masses while preserving the number of degrees of freedom can only describe the spontaneous breaking of the gauge symmetry in the Coulomb branch. Hence, both $(G+B)_{IJ}$  and $Y_{Iu}$ may be interpreted as Wilson lines along the compact direction $X^I$ of the $(10-d)+16$ Cartan $U(1)$'s.

The marginal deformations admits periodicity and symmetry properties. Using
\be
e^{-{\pi\over \Im \tau}(\tilde m_I+n_I\tau)(G+B)_{IJ}(\tilde m_J+n_J\bar \tau)}
= e^{-2i\pi n_IB_{IJ}\tilde m_J}~,
\ee
one concludes that 
\be
B_{IJ}~~ \longrightarrow ~~ B_{IJ}+\delta B_{IJ}~,~~\quad \where\quad ~~\delta B_{IJ}\in \Z~,~~I,J\in\{d,\dots,9\}~,
\ee
is a periodicity of $\Gamma_{10-d,26-d}\colvec[0.7]{\gamma \\ \delta}$. Moreover, one can check that
\be
\begin{aligned}
&(Y_{Ku}, B_{IJ})~~ \longrightarrow ~~ \Big(Y_{Ku}+\delta Y_{Ku}\, ,B_{IJ}-\half \big(\delta Y_{Iu} Y_{Ju}-Y_{Iu} \delta Y_{Ju}\big)\Big)\; , \\
 \where\quad ~~&\delta Y_{Ku}\in 2\Z~, ~~K\in\{d,\dots,9\}~,~~u\in\{1,\dots, 16\}~,\esp
\end{aligned}
\ee
is another symmetry. Finally, for any given $I\in\{d,\dots,9\}$, the inversion $\vec Y_{I}\to -\vec Y_{I}$ leaves $\Gamma_{10-d,26-d}\colvec[0.7]{\gamma \\ \delta}$ invariant. 


\subsection{\boldmath $\pmb{\Z}_2$-orbifold actions on 16  $\psi^i$'s}
\label{16}

The link between Wilson-line backgrounds and free orbifold actions can be illustrated by models constructed in the previous section, which involve the generators $G_1^\f,\dots, G_5^\f$. 


\paragraph{\boldmath \em $\pmb{\Z}_2$ generated by $G_1^{\rm f}$:}

The first example we consider is that of the $\Spin$ theory compactified on a circle $S^1(R_9)$ of radius $R_9$.\footnote{We measure it in string units, \ie we set the string tension $\alpha'$ to be 1.}  The moduli space is spanned by $R_9\equiv \sqrt{G_{99}}$ and the Wilson lines $Y_{9u}$, $u\in\{1,\dots, 16\}$. The latter can be split into a background value $\hat Y_{9u}$ and a continuous deformation $Y'_{9u}$ as follows,
  \begin{align}
 \vec Y_{9}&=\vec Y'_{9}+\vec{\hat Y}_{9}~,\label{back1}\\
\where ~~\quad \vec {\hat Y}_9&=\big(\,0~,~\,0~,~\,0~,~\,0~,~\,0~,~\,0~,~\,0~,~\,0~,\mbox{$-{1\over 2}$},\mbox{$-{1\over 2}$},\mbox{$-{1\over 2}$},\mbox{$-{1\over 2}$},\mbox{$-{1\over 2}$},\mbox{$-{1\over 2}$},\mbox{$-{1\over 2}$},\mbox{$-{1\over 2}$}~ \big)~ .\nonumber 
\end{align}
Inserting this decomposition of $\vec Y_9$ into \Eq{block} for $d=9$, and redefining
\be
n_9=2(\ell_9+h_1)~,~~\quad \tilde m_9=2(\tilde k_9+g_1)~,~~\quad \where\quad h_1,g_1\in\Big\{0,\half\Big\}~,
\ee
in order to divide the discrete sums over $n_9,\tilde m_9$ into sums over $\ell_9,\tilde k_9$ and $h_1,g_1$, one obtains 
\begin{align}
\Gamma_{1,17}\colvec[0.7]{\gamma \\ \delta}\big(R_9^2,\vec Y_9'+\vec {\hat Y}_9\big)=&\; {2R_9\over \sqrt{\Im \tau}}\,\half \sum_{h_1,g_1}\sum_{\ell_9,\tilde k_9}e^{-{\pi (2R_9)^2\over \Im \tau}\left|\tilde k_9+g_1+(\ell_9+h_1)\tau\right|^2}\, e^{4i\pi (\ell_9+h_1)\vec Y'_{9} \cdot \left(\delta \vec \II_{16}-(\tilde k_9+g_1)\vec Y'_{9}\right)}\nonumber  \\
&\times \prod_{u=1}^{8} \vartheta\colvec[0.7]{\gamma -2(\ell_9+h_1)Y'_{9u}\\ \delta-2(\tilde k_9+g_1)Y'_{9u}} \prod_{v=9}^{16} \Big(e^{4i\pi (\ell_9+h_1)g_1Y_{9v}'}\, \vartheta\colvec[0.7]{\gamma+h_1 -2(\ell_9+h_1)Y'_{9v}\\ \delta+g_1-2(\tilde k_9+g_1)Y'_{9v}}\Big)\, . 
\label{WLex}
\end{align}
At the new origin of the Wilson line moduli space, $\vec Y'_9=\vec 0$, this leads to  
\begin{align}
\half\sum_{\gamma,\delta}\Gamma_{1,17}\colvec[0.7]{\gamma \\ \delta}\big(R_9^2,\vec{\hat Y}_9\big)&=\half \sum_{h_1,g_1}{2R_9\over \sqrt{\Im \tau}}\sum_{\ell_9,\tilde k_9}e^{-{\pi (2R_9)^2\over \Im \tau}\left|\tilde k_9+g_1+(\ell_9+h_1)\tau\right|^2}\,\half\sum_{\gamma,\delta} \vartheta\colvec[0.7]{\gamma \\ \delta}^8\, \vartheta\colvec[0.7]{\gamma+h_1\\ \delta+g_1}^8\nonumber \\
&\equiv \half \sum_{h_1,g_1}\Gamma_{1,1}\colvec[0.7]{h_1 \\ g_1}(4R_9^2)\, \Gamma^\Spin\colvec[0.7]{h_1 \\ g_1}~ .
\end{align}
where we recognize the lattice of zero modes of the circle of double radius $2R_9$ (see \Eq{ga}), coupled to the $\Spin$ root lattice modded by $G_1$ (see \Eq{la1}). Hence, the Wilson-line background~(\ref{back1}) of the $\Spin$ theory compactified on $S^1(R_9)$ is equivalent to the background of the $\Spin$ theory compactified on $S^1(2R_9)$ and orbifolded by $\Z_2$ generated by $G_1^\f$.   

More generally, in the $\Spin$ theory compactified toroidally on $T^{10-d}$, one can always derive from \Eq{block}   by following the same steps the continuous Wilson-line deformations around any free orbifold model  viewed as  as a Wilson-line background. Hence, from now on,  we will concentrate on the orbifold descriptions of specific backgrounds and will not implement these continuous Wilson-line deformations.  


\paragraph{\boldmath \em $\pmb{\Z}^4_2$ generated by $G_1^{\rm f},\dots,G_4^{\rm f}$:}

From the previous example, it seems natural to expect that orbifold models obtained by combining the actions of several $\Z_2$ generators $G_\alpha^\f$ in lower dimension may be described by Wilson-line backgrounds where all components $Y_{Iu}$ take values in $\{0,-\half\}$. For instance, in dimension $d=6$, the background naturally generalizing that given in \Eq{back1} may be
\be
 \begin{aligned}
\vec {\hat Y}_9&=\big(\,0~,~\,0~,~\,0~,~\,0~,~\,0~,~\,0~,~\,0~,~\,0~,\mbox{$-{1\over 2}$},\mbox{$-{1\over 2}$},\mbox{$-{1\over 2}$},\mbox{$-{1\over 2}$},\mbox{$-{1\over 2}$},\mbox{$-{1\over 2}$},\mbox{$-{1\over 2}$},\mbox{$-{1\over 2}$}~ \big) ~,\\
\vec {\hat Y}_8&=\big(\,0~,~\,0~,~\,0~,~\,0~,\mbox{$-{1\over 2}$},\mbox{$-{1\over 2}$},\mbox{$-{1\over 2}$},\mbox{$-{1\over 2}$},~\,0~,~\,0~,~\,0~,~\,0~,\mbox{$-{1\over 2}$},\mbox{$-{1\over 2}$},\mbox{$-{1\over 2}$},\mbox{$-{1\over 2}$}~ \big) ~, \\
\vec {\hat Y}_7&=\big(\,0~,~\,0~,\mbox{$-{1\over 2}$},\mbox{$-{1\over 2}$},~\,0~,~\,0~,\mbox{$-{1\over 2}$},\mbox{$-{1\over 2}$},~\,0~,~\,0~,\mbox{$-{1\over 2}$},\mbox{$-{1\over 2}$},~\,0~,~\,0~,\mbox{$-{1\over 2}$},\mbox{$-{1\over 2}$}~ \big)~,\\ 
\vec {\hat Y}_6&=\big(\,0~,\mbox{$-{1\over 2}$},~\,0~,\mbox{$-{1\over 2}$},~\,0~,\mbox{$-{1\over 2}$},~\,0~,\mbox{$-{1\over 2}$},~\,0~,\mbox{$-{1\over 2}$},~\,0~,\mbox{$-{1\over 2}$},~\,0~,\mbox{$-{1\over 2}$},~\,0~,\mbox{$-{1\over 2}$}~ \big)~.
\end{aligned}
\label{b1}
\ee
However, for order 2 generators, background values equal to $\mbox{$+{1\over 2}$}$ should play similar roles. Notice, though, that from the symmetry properties given at the end of \Sect{wlT}, the choice of background for any $Y_{Iu}$ may be restricted to lie in the range $(-1,1]$, provided that the antisymmetric tensor is kept arbitrary. Hence, background values equal to $\mbox{$-{1\over 2}$}$ and $\mbox{$+{1\over 2}$}$ may not be totally equivalent. To understand the difference,\footnote{One may compare similarly background values equal to 0 and 1.} let us consider an alternative example of Wilson-line vacuum expectation values  in six dimensions  given by 
\be
 \begin{aligned}
\vec {\check Y}_9&=\big(\,0~,~\,0~,~\,0~,~\,0~,~\,0~,~\,0~,~\,0~,~\,0~,\mbox{$-{1\over 2}$},\mbox{$-{1\over 2}$},\mbox{$-{1\over 2}$},\mbox{$-{1\over 2}$},\mbox{$+{1\over 2}$},\mbox{$+{1\over 2}$},\mbox{$+{1\over 2}$},\mbox{$+{1\over 2}$}~ \big)~, \\
\vec {\check Y}_8&=\big(\,0~,~\,0~,~\,0~,~\,0~,\mbox{$-{1\over 2}$},\mbox{$-{1\over 2}$},\mbox{$+{1\over 2}$},\mbox{$+{1\over 2}$},~\,0~,~\,0~,~\,0~,~\,0~,\mbox{$+{1\over 2}$},\mbox{$+{1\over 2}$},\mbox{$-{1\over 2}$},\mbox{$-{1\over 2}$}~ \big)~,  \\
\vec {\check Y}_7&=\big(\,0~,~\,0~,\mbox{$-{1\over 2}$},\mbox{$+{1\over 2}$},~\,0~,~\,0~,\mbox{$+{1\over 2}$},\mbox{$-{1\over 2}$},~\,0~,~\,0~,\mbox{$-{1\over 2}$},\mbox{$+{1\over 2}$},~\,0~,~\,0~,\mbox{$+{1\over 2}$},\mbox{$-{1\over 2}$}~ \big)~,\\ 
\vec {\check Y}_6&=\big(\,0~,\mbox{$-{1\over 2}$},~\,0~,\mbox{$+{1\over 2}$},~\,0~,\mbox{$-{1\over 2}$},~\,0~,\mbox{$+{1\over 2}$},~\,0~,\mbox{$-{1\over 2}$},~\,0~,\mbox{$+{1\over 2}$},~\,0~,\mbox{$-{1\over 2}$},~\,0~,\mbox{$+{1\over 2}$}~ \big)~.
\end{aligned}
\label{b2}
\ee

For the backgrounds $\vec Y\equiv\vec {\check Y}$ and $\vec Y\equiv\vec {\hat Y}$, \Eq{block} turns out to yield the final expressions\footnote{Details of the derivation are postponed to the next paragraph, when we discuss the case of a $\Z_2^5$ orbifold in 5 dimensions.}
\begin{align}
\half\sum_{\gamma,\delta}\Gamma_{4,20}\colvec[0.7]{\gamma \\ \delta}\big(G,B,\vec {\check Y}\big)&= {1\over 2^4} \sum_{\substack{h_1,\dots,h_4 \\ g_1,\dots, g_4}} \Gamma_{4,4}\colvec[0.7]{h_1,\dots,h_4 \\ g_1,\dots,g_4}(4G,4B)\, \Gamma^\Spin\colvec[0.7]{h_1,\dots,h_4 \\ g_1,\dots,g_4} \; ,\label{G4}
\\
\half\sum_{\gamma,\delta}\Gamma_{4,20}\colvec[0.7]{\gamma \\ \delta}\big(G,B,\vec {\hat Y}\big)&= {1\over 2^4} \sum_{\substack{h_1,\dots,h_4 \\ g_1,\dots, g_4}} \Gamma_{4,4}\colvec[0.7]{h_1,\dots,h_4 \\ g_1,\dots,g_4}(4G,4B)\, \Gamma^\Spin\colvec[0.7]{h_1,\dots,h_4 \\ g_1,\dots,g_4} \, \T\colvec[0.7]{h_1,\dots,h_4 \\ g_1,\dots,g_4}~ .\nonumber 
\end{align}
Therefore, the  background $(G+B)_{IJ}$, $\vec {\check Y}_I$ of the $\Spin$ theory compactified on $T^5$ is equivalent to the background $4(G+B)_{IJ}$, $\vec Y_I=\vec 0$ of the $\Spin$ theory compactified on $T^5$ and modded by $\Z_2^4$  generated by  $G_1^\f,\dots,G_4^\f$. Moreover, the  background $(G+B)_{IJ}$, $\vec {\hat Y}_I$ is similar, up to a discrete torsion~\cite{torsion-1,torsion-2},
\be
\T\colvec[0.7]{h_1,\dots,h_4 \\ g_1,\dots,g_4} = (-1)^{4\sum_{\alpha \neq \beta}h_\alpha g_\beta}~ ,
\ee
which is a set of signs that modifies the projections onto $G_\alpha^\f$-invariant states, $\alpha=1,\dots,4$. If, by construction, we already know that both lattices in \Eq{G4} yield modular invariant one-loop partition functions, it can  be checked \emph{a posteriori} that the presence or not of such discrete torsions does not alter modular invariance. Indeed, under the reshuffling of the conformal blocks induced by modular transformations,
\be
\begin{aligned}
\tau\to \tau+1\quad &\Longrightarrow \quad (h_{\alpha},g_{\alpha})\to (h_{\alpha},g_{\alpha}+h_{\alpha})~,~~\quad \alpha\in\{1,\dots,4\}~,\\
\tau\to-{1\over  \tau}\quad &\Longrightarrow \quad (h_{\alpha},g_{\alpha})\to (g_{\alpha},-h_{\alpha})~,
\end{aligned}
\ee
any individual sign $(-1)^{4h_{\alpha_0}g_{\beta_0}+4h_{\beta_0}g_{\alpha_0}}$ for given $\alpha_0,\beta_0\in\{1,\dots,4\}$, $\alpha_0\neq \beta_0$, is  invariant. Moreover, it is invariant under the shift by 1 of any of the $h_{\alpha_0}, h_{\beta_0}, g_{\alpha_0},g_{\beta_0}$. Hence, such a sign can be introduced or removed from the partition function at will. In fact, it is only in dimension $d=9$ that such discrete torsions that distinguish Wilson-line expectation values $-\half$ and $+\half$ do not exist.

As explained in \Sect{redrank}, in order to find the generic gauge symmetry induced by the fermions $\psi^i$, it is enough to restrict in the Hamiltonian form of the $T^4$ lattice to the untwisted states at zero momentum and winding numbers. For both backgrounds, this yields  $SO(2)^{16}$, as the torsion is only non-trivial in the twisted sectors. Indeed, one obtains using \Eq{g1g2g3-2} that
\be
\begin{aligned}
 {1\over 2^4}\sum_{g_1,\dots ,g_4} \Gamma_{4,4}\colvec[0.7]{0,\dots,0 \\ g_1,\dots,g_4}\Big|_{\substack{m_6=\dots=m_9=0 \\ n_6=\dots= n_9=0}}\,&\frac{\Gamma^\Spin\colvec[0.7]{0,\dots,0 \\ g_1,\dots,g_4}}{\eta^{16}}\, \T\colvec[0.7]{0,\dots,0 \\ g_1,\dots,g_4}\\
&= {1\over 2^4}\sum_{g_1,\dots ,g_4} \frac{\Gamma^\Spin\colvec[0.7]{0,\dots,0 \\ g_1,\dots,g_4}}{\eta^{16}}=O_2^{16} + \text{massive}~ .\esp
\end{aligned}
\label{so216}
\ee
Finally,  note that it is also clear from the orbifold point of view that both lattices  in \Eq{G4} admit identical continuous Wilson-line deformations. This follows from the fact that the orbifold actions are free, implying all moduli, which are massless, to belong to the untwisted sector. However, the latter is not affected by the presence or not of the  discrete torsion $\T$ which modifies projections only in the twisted sectors. Moreover, let us mention that away from the isolated backgrounds involving only expectations values for $Y_{Iu}$ in $\{-\half,0,\half,1\}$, the gauge symmetry generated by the fermions $\psi^i$ is generic and contains unitary factors. 


\paragraph{\boldmath \em $\pmb{\Z}^5_2$ generated by $G_1^{\rm f},\dots,G_5^{\rm f}$:}

We know that in the free-orbifold model associated with the generators $G_1^\f,\dots, G_5^\f$ in five dimensions, the gauge symmetry generated by the fermions $\psi^i$ is trivial, $SO(1)^{32}$, implying no Wilson-line deformations $Y_{Iu}$ to exist. In the following, we show that a  generalized version of formula~(\ref{block})  can nevertheless be used to construct consistent orbifold models of reduced rank. 

In arbitrary dimension $d$, our starting point may be  \Eq{block} rewritten as 
\be
\begin{aligned}
\Gamma_{10-d,26-d}&=\half\sum_{\gamma,\delta}\Gamma_{10-d,26-d}\colvec[0.7]{\gamma \\ \delta}\;,~~\quad \where\\
\Gamma_{10-d,26-d}\colvec[0.7]{\gamma \\ \delta}(G,B,\vec y)&={\sqrt{\det G}\over (\Im \tau)^{10-d\over 2}}\sum_{\substack{n_d,\dots,n_9 \\ \tilde m_d,\dots, \tilde m_9}} e^{-{\pi\over \Im \tau}(\tilde m_I+n_I\tau)(G+B)_{IJ}(\tilde m_J+n_J\bar \tau)} \\
&\hspace{3cm} \times e^{i{\pi\over 2} n_I\vec y_{I} \cdot \left(2\delta \vec \II_{32}-\tilde m_J\vec y_{J}\right)} \, \prod_{i=1}^{32} \vartheta\colvec[0.7]{\gamma -n_Ky_{Ki}\\ \delta-\tilde m_Ly_{Li}}^{\half}~ .
\end{aligned}
\label{blocky}
\ee
In this expression, $\vec y_{I}$ and $\vec \II_{32}$ are 32-vectors with entries $y_{Ii}$ or 1, respectively. Hence, we replace the $(10-d)\times 16$ continuous Wilson lines $Y_{Iu}$ by  $(10-d)\times 32$ deformations $y_{Ii}$ that may or may not be interpreted as moduli, depending on the case at hand. Clearly, when 
\be
y_{I,2u-1}\equiv y_{I,2u}\equiv Y_{Iu}~,~~\quad I\in\{d,\dots, 9\}~,~~u\in\{1,\dots, 16\}~,
\ee
we recover the initial Narain lattice~(\ref{block}), which leads to rank $10-d+16$ gauge groups generated by the bosonic side of the $\Spin$ heterotic string. In all other cases, \Eq{blocky} can be used to derive consistent free orbifold models of reduced ranks. Indeed, all properties of $\Gamma_{10-d,26-d}\colvec[0.7]{\gamma \\ \delta}$ regarding modular transformations remain valid whether it is deformed by $Y_{Iu}$'s or $y_{Ii}$'s. 

In dimension $d=5$, let us consider two possible deformations. The first one, denoted  $\vec{\hat y}_I$, is equivalent to the background given in \Eq{b1} supplemented by deformations associated with the extra dimension 5,  
\be
\begin{aligned}
\hat y_{I,2u-1}&= \hat y_{I,2u}=\hat  Y_{Iu}~,~~&& I\in\{6,\dots, 9\}~,~~u\in\{1,\dots, 16\}~,\\
\hat y_{5,2u-1}&=0~,~~\hat y_{5,2u}=-\half~.
\end{aligned}
\ee
The second one, $\vec{\check y}_I$, refines similarly that given in \Eq{b2},
\be
\begin{aligned}
\check y_{I,2u-1}&= \check y_{I,2u}=\check  Y_{Iu}~,~~\quad  I\in\{6,\dots, 9\}~,~~u\in\{1,\dots, 16\}~,\\
\check y_{5,4u-3}&=0~,~~\check y_{5,4u-2}=-\half~,~~\check y_{5,4u-1}=0~,~~\check y_{5,4u}=+\half~,~~\quad u\in\{1,\dots,8\}~.
\end{aligned}
\ee
Taking $\vec y=\vec {\hat y}$ in  \Eq{blocky} and redefining
\be
n_I=2(\ell_I+h_{10-I})~,~~~ \tilde m_I=2(\tilde k_I+g_{10-I})~,~~~ \where~~~ h_{10-I},g_{10-I}\in\Big\{0,\half\Big\}~,~~I\in\{5,\dots,9\}~,
\ee
we obtain
\be
\begin{aligned}
\half\sum_{\gamma,\delta}\Gamma_{5,21}\colvec[0.7]{\gamma \\ \delta}\big(G,B,\vec {\hat y}\,\big)&= {1\over 2^5} \sum_{\substack{h_1,\dots,h_5 \\ g_1,\dots, g_5}} \Gamma_{5,5}\colvec[0.7]{h_1,\dots,h_5 \\ g_1,\dots,g_5}(4G,4B)\, \Gamma^\Spin\colvec[0.7]{h_1,\dots,h_5 \\ g_1,\dots,g_5} \, \T\colvec[0.7]{h_1,\dots,h_5 \\ g_1,\dots,g_5}~ ,
\end{aligned}
\label{G51}
\ee
where the discrete torsion 
\be
\T\colvec[0.7]{h_1,\dots,h_5 \\ g_1,\dots,g_5} = (-1)^{4\sum_{\alpha \neq \beta}h_\alpha g_\beta} 
\ee
arises from the phase $e^{i{\pi\over 2} n_I\vec {\hat y}_{I} \cdot \left(2\delta \vec \II_{32}-\tilde m_J\vec {\hat y}_{J}\right)}$. 
For the second deformation, $\vec y=\vec {\check y}$, notice that we have chosen assignments of $\pm \half$ deformations such that 
\be
\vec{\check y}_I\cdot \vec\II_{32}=0~,~~\quad \vec{\check y}_I\cdot \vec{\check y}_J=4\delta_{IJ}~,~~\quad I,J\in\{5,\dots,9\}~,
\ee
which implies $e^{i{\pi\over 2} n_I\vec {\check y}_{I} \cdot \left(2\delta \vec \II_{32}-\tilde m_J\vec {\check y}_{J}\right)}=1$. Moreover, observing that $\vec{\check y}_I=\vec{\hat y}_I+\vec \upsilon_I$ where \mbox{$\vec \upsilon_I\cdot \vec{\hat y}_J\in4\Z$}, $I,J\in\{5,\dots,9\}$, one can show that
\be
\prod_{i=1}^{32} \vartheta\colvec[0.7]{\gamma -n_K\check y_{Ki}\\ \delta-\tilde m_L\check y_{Li}}^{\half}=\prod_{i=1}^{32} \vartheta\colvec[0.7]{\gamma -n_K\hat y_{Ki}\\ \delta-\tilde m_L\hat y_{Li}}^{\half}~.
\ee
Hence, we conclude that 
\be
\begin{aligned}
\half\sum_{\gamma,\delta}\Gamma_{5,21}\colvec[0.7]{\gamma \\ \delta}\big(G,B,\vec {\check y}\,\big)&= {1\over 2^5} \sum_{\substack{h_1,\dots,h_5 \\ g_1,\dots, g_5}} \Gamma_{5,5}\colvec[0.7]{h_1,\dots,h_5 \\ g_1,\dots,g_5}(4G,4B)\, \Gamma^\Spin\colvec[0.7]{h_1,\dots,h_5 \\ g_1,\dots,g_5} ~ .
\end{aligned}
\label{G52}
\ee

The lattices in \Eq{G51} and~(\ref{G52}) are those encountered in the $\Spin$ theory on $T^5$, with internal metric and antisymmetric tensors $4G,4B$, and orbifolded by $\Z_2^5$ generated by $G_1^\f,\dots,G_5^\f$, with or without discrete torsion.  As in the $\Z_2^4$ case described before,  they have identical untwisted sectors. In particular, the ``gauge symmetries'' generated by the fermions $\psi^i$ are in both cases $SO(1)^{32}$. The Coulomb branch of such a trivial group being zero-dimensional, the deformations $\vec{\hat y}$ or  $\vec{\check y}$ cannot be made continuous. Let us stress again that the essence of \Eq{blocky} is its good transformation properties under the modular group, irrespectively of the fact that $\vec y$ are expectations values of moduli fields or simply discrete deformations. It is  because none of the pairs ${\hat y}_{5,2u-1},{\hat y}_{5,2u}$ (or $\check y_{5,2u-1},\check y_{5,2u}$) take equal values for any \mbox{$u\in\{1,\dots,16\}$} that the rank is reduced by 16 units.  
In the next subsection, we describe another example where only part but not all of the pairs $y_{I,2u-1},y_{I,2u}$ cannot be combined in Wilson-line moduli expectation values.


\subsection{\boldmath $\pmb{\Z}_2$-orbifold actions on 8  $\psi^i$'s}
\label{08}

So far, we have only considered orbifold generators that flip the signs of 16 fermions $\psi^i$.  This is because we started our discussion in \Sect{d10} in 10 dimensions, for which it can be easily seen that only generators acting on a multiple of 8 fermions $\psi^i$ may be considered, as imposed by modular invariance. In the following, we  analyze Wilson-line backgrounds equivalent to orbifold models involving generators acting on 8 fermions. This will be the opportunity to see that unlike the case where all $\Z_2$ free generators act on 16 $\psi^i$'s, $\Z_2$ free generators acting on less than 16 fermions must act in most cases ``asymmetrically'' on the lattice of zero modes of the internal torus. We will consider in particular an example of model with rank reduced by 8 units and involving $SO(3)$ gauge group factors. 


\paragraph{\boldmath \em $\pmb{\Z}_2$ free generator $\tilde G^{\rm f}$ acting on 8 $\psi^i$'s:}

Our starting point is the $\Spin$ theory compactified on $S^1(R_9)$ with Wilson-line background 
\be
\vec {\tilde Y}_9=\big(\,0~,~\,0~,~\,0~,~\,0~,~\,0~,~\,0~,~\,0~,~\,0~,~\,0~,~\,0~,~\,0~,~\,0~,\mbox{$-{1\over 2}$},\mbox{$-{1\over 2}$},\mbox{$-{1\over 2}$},\mbox{$-{1\over 2}$}~ \big)~ .
\ee
Applying \Eq{block} for $d=9$ at $\vec {Y}_9=\vec {\tilde Y}_9$, and redefining 
\be
n_9=2(\ell_9+\tilde h)~,~~\quad \tilde m_9=2(\tilde k_9+\tilde g)~,~~\quad \where\quad \tilde h,\tilde g\in\Big\{0,\half\Big\}~,
\ee
we obtain
\be
\half\sum_{\gamma,\delta}\Gamma_{1,17}\colvec[0.7]{\gamma \\ \delta}\big(R_9^2,\vec{\tilde Y}_9\big)= \half \sum_{\tilde h,\tilde g}\Gamma_{1,1}\colvec[0.7]{\tilde h \\ \tilde g}(4R_9^2)\,  \Gamma^\Spin\colvec[0.7]{\tilde h \\ \tilde g}\, (-1)^{4\tilde h\tilde g}~ ,
\label{orb1}
\ee
where we denote  from now on
\be
\Gamma^\Spin\colvec[0.7]{\tilde h \\ \tilde g}\equiv\half\sum_{\gamma,\delta} \vartheta\colvec[0.7]{\gamma \\ \delta}^{12}\, \vartheta\colvec[0.7]{\gamma+\tilde h\\ \delta+\tilde g}^4~,
\label{defbis}
\ee
which differs from the  definition used in \Eq{la1}. Notice the presence of a sign $(-1)^{4\tilde h\tilde g}$ which is a remnant  of the phase that dresses the product of $\vartheta$ functions in \Eq{block}. This sign is not a discrete torsion, as it is required by modular invariance. Indeed, it makes $(-1)^{4\tilde h\tilde g}\Gamma^\Spin\colvec[0.7]{\tilde h \\ \tilde g}$ transform suitably under modular transformations for the contributions $(\tilde h,\tilde g)$ of the partition function to be reshuffled,  
\be
\begin{aligned}
\tau\to \tau+1\quad &\Longrightarrow \quad (\tilde h,\tilde g)\to (\tilde h,\tilde g+\tilde h)~,\\
\tau\to-{1\over  \tau}\quad &\Longrightarrow \quad (\tilde h,\tilde g)\to (\tilde g,-\tilde h)~,
\end{aligned}
\ee
while preserving the definitions of $\tilde h,\tilde g$ modulo 1. 

\Eq{orb1} is telling us that the background $\vec{\tilde y}$ of the $\Spin$ theory compactified on $S^1(R_9)$ admits a description in terms of a $\Z_2$-orbifold action on the $\Spin$ theory compactified on $S^1(2R_9)$. The associated generator $\tilde G^\f$ acts as 
\be
\tilde G^\f = \tilde G \otimes \big( X^9\rightarrow X^9+\pi \big) ~,
\ee
where $\tilde G$ flips the signs of 8 fermions $\psi^i$, 
\be
\tilde G : ++++++++++++++++++++++++-------- ~ .
\ee
Moreover, notice that changing $\Gamma_{1,1}\colvec[0.7]{\tilde h \\ \tilde g}(4R_9^2)\to 2R_9/ \sqrt{\Im \tau}$ in the r.h.s. of \Eq{orb1} does not alter the modular transformations properties. Therefore, we may consider the action of $\tilde G$ alone, \ie on the heterotic strings in ten dimensions.  
In fact, if the Wilson-line background we started with had not led to a generator acting on a multiple of 8 fermions, the phase dressing the $\vartheta$ functions in \Eq{block} (or \Eq{blocky}) would remain dependant on $\ell_9,\tilde k_9$. In such cases, it would not be consistent to consider a version of the orbifold action in 10 dimensions. 


\paragraph{\boldmath \em Action of $\tilde G$ in ten dimensions:}

Before illustrating in the next paragraph the non-trivial use of free generators acting on 8 fermions $\psi^i$, let us show that unlike  $G_1,\dots G_5$, the generator $\tilde G$ acts somehow trivially in ten dimensions.\footnote{We thank Carlo Angelantonj for private communication on this statement.} 

To derive its effect on the $\Spin$ and $E_8\times E_8$ theories, we apply the method utilized in \Sect{h8} and~\ref{8h}, which is based on the decomposition formulas~(\ref{decomp1}) and transformations rules~(\ref{Gtr}),~(\ref{Gtwist}). The action of $\tilde G$ on the $\Spin$ lattice reads
\begin{align}
\half \sum_{\tilde h,\tilde g}{\Gamma^\Spin \colvec[0.7]{\tilde h\\ \tilde g}\over \eta^{16}} \, (-1)^{4\tilde h\tilde g}&= \half \sum_{\tilde g}\Big\{O_{24}O_ {8} + V_{24}(-1)^{2\tilde g}V_{8} + S_{24}S_{8} + C_{24}(-1)^{2\tilde g}C_{8} \nonumber\\
&\hspace{0.05cm}+ (-1)^{2\tilde g}\Big[O_{24}S_ {8} + V_{24}(-1)^{2\tilde g}C_{8} + S_{24}O_{8} + C_{24}(-1)^{2\tilde g}V_{8}\Big]\Big\}\nonumber\\
&=O_{24}O_ {8} +S_{24}S_ {8} +V_{24}C_ {8} +C_{24}V_ {8} \\
&=O_{32}'+S_{32}'~,\nonumber 
\end{align}
where we have used the triality symmetry $V_8\leftrightarrow C_8$ to obtain the last equality. Similarly, $\tilde G$ acting on the $E_8\times E_8$ theory yields
\begin{align}
\half \sum_{\tilde h,\tilde g}{\Gamma^{E_8\times E_8} \colvec[0.7]{\tilde h\\ \tilde g}\over \eta^{16}} \, (-1)^{4\tilde h\tilde g}&= \half \sum_{\tilde g}(O_{16}+S_{16}) \Big\{O_{8}O_ {8} + V_{8}(-1)^{2\tilde g}V_{8} + S_{8}S_{8} + C_{8}(-1)^{2\tilde g}C_{8} \nonumber\\
&\hspace{2cm}+ (-1)^{2\tilde g}\Big[O_{8}S_ {8} + V_{8}(-1)^{2\tilde g}C_{8} + S_{8}O_{8} + C_{8}(-1)^{2\tilde g}V_{8}\Big]\Big\}\nonumber\\
&=(O_{16}+S_{16}) (O_{8}O_ {8} +S_{8}S_ {8} +V_{8}C_ {8} +C_{8}V_ {8}) \\
&=(O_{16}+S_{16})(O'_{16}+S'_{16})~,\nonumber 
\end{align}
 where we use the triality $V_8\leftrightarrow C_8$ for the characters appearing in last positions in the monomials. From the above equations, we see that $\tilde G$ actually maps both heterotic strings into themselves. 
 

\paragraph{\boldmath \em Rank reduced by 8 units in six dimensions:} 

Let us turn to the case of a background in six dimensions that is  equivalent to a $\Z_2^4$ free orbifold model, and those gauge symmetry has a rank reduced by 8 units. We will denote the four generators derived from the Wilson-line approach $G_1^{\f\prime},G_2^{\f\prime},G_3^{\f\prime},\tilde G_4^\f$. We want them to act on the $\Spin$ lattice in order to reduce the dimension and rank of the gauge symmetry generated by the fermions $\psi^i$'s, and also to operate on $T^4$ for their global actions to be free. In that case, extra gauge bosons will not enhance  the gauge group back. To be specific, we look for generators whose structures are of the forms
\be
\begin{aligned}
G_{10-I}^{\f\prime} &= G_{10-I} \otimes \big( \mbox{free action on $T^4$} \big) \, ,~~\quad I\in\{7,8,9\}~,\\
\tilde G_{4}^\f &= \tilde G_{4} \otimes \big(  \mbox{free action on $T^4$} \big)\, , 
\end{aligned}
\ee
where $G_1,G_2,G_3$ are defined in \Sect{23} while $\tilde G_4$ flips only  8 fermions. The latter act on the $\psi^i$'s by multiplying them with signs listed below,
 \be
\begin{aligned}
G_1 :&\ ++++++++++++++++---------------- ~ , \\
G_2 :&\ ++++++++--------++++++++-------- ~ , \\
G_3 :&\ ++++----++++----++++----++++---- ~ , \\
\tilde G_4 :&\ +++-+++-\hspace{0.023cm}+++-+++-\hspace{0.023cm}+++-+++-\hspace{0.023cm}+++-+++- ~ .
\end{aligned}
\ee 
In the following, our task is to determine the associated free actions on $T^4$. 

Let us consider the deformations of the $\Spin$ theory compactified on $T^4$ 
\be
\begin{aligned}
\tilde y_{I,2u-1}&= \tilde y_{I,2u}=\tilde   Y_{Iu}~,~~\quad  I\in\{7,8, 9\}~,~~u\in\{1,\dots, 16\}~,\\
\tilde y_{6,4u-3}&=\tilde y_{6,4u-2}=\tilde Y_{6,2u-1}=0~,~~\tilde y_{6,4u-1}=0~,~~\tilde y_{6,4u}=-\half~,~~\quad u\in\{1,\dots,8\}~,
\end{aligned}
\label{deforma}
\ee
where $\vec{\tilde Y}_6,\vec{\tilde Y}_7,\vec{\tilde Y}_8$ are identical to $\vec{\hat Y}_6, \vec{\hat Y}_7, \vec{\hat Y}_8$ defined in \Eq{b1}.  Notice that  all $\tilde Y_{I,2u-1}$, $I\in\{6,\dots,9\}$, $u\in\{1,\dots,8\}$,  are well defined (\ie equal to $\tilde y_{I,4u-3}=\tilde y_{I,4u-2}$), and can therefore be deformed continuously. They are \emph{bona fide} vacuum expectation values of Wilson-line moduli. On the contrary, because $\tilde y_{6,4u-1}\neq \tilde y_{6,4u}$, $u\in\{1,\dots,8\}$, these deformations are not related to massless scalar fields, which  implies  the rank to be reduced by 8 units. As a result, all of the  $\tilde y_{I,4u-1},\tilde y_{I,4u}$, $I\in\{6,\dots,9\}$, $u\in\{1,\dots,8\}$, are actually discrete parameters. 

Applying \Eq{blocky} for $d=6$ at $\vec y=\vec{\tilde y}$ and redefining as usual  
\be
n_I=2(\ell_I+h_{10-I})~,~~~ \tilde m_I=2(\tilde k_I+g_{10-I})~,~~~ \where~~~ h_{10-I},g_{10-I}\in\Big\{0,\half\Big\}~,~~I\in\{6,\dots,9\}~,
\ee
we obtain a result that can be interpreted as a lattice orbifolded by a $\Z_2^4$ group, and for background torus moduli $4G_{IJ}, 4B_{IJ}$, 
\begin{align}
\half\sum_{\gamma,\delta}\Gamma_{4,20}\colvec[0.7]{\gamma \\ \delta}\big(G,B,\vec {\tilde y}\,\big)= {1\over 2^4} \sum_{\substack{h_1,\dots,h_4 \\ g_1,\dots, g_4}}  \Gamma_{4,4}&\colvec[0.7]{\goh_1,\dots,\goh_4;\goh'_1,\dots,\goh'_4  \\ 
\gog_1,\dots,\gog_4;\gog'_1,\dots,\gog'_4}(4G,4B)\, \Gamma^\Spin\colvec[0.7]{h_1,\dots,h_4 \\ g_1,\dots,g_4}\nonumber \\ 
\times \,\Phi &\colvec[0.7]{h_1,\dots,h_4 \\ g_1,\dots,g_4}\;\T\colvec[0.7]{h_1,h_2,h_3 \\ g_1,g_2,g_3}~ ,\label{G6}
\end{align}
where $\T$, $\Phi$, $\Gamma_{4,4}$ and $h_\alpha',g_\alpha'$, $\alpha\in\{1,\dots,4\}$,  are defined as follows:\footnote{It is understood that the last column $\colvec[0.7]{h_4 \\ g_4}$ labelling the orbifold action on the $\Spin$ lattice refers to the definition~(\ref{defbis}). } 

$\bullet$ Firstly, we have 
\be
\T\colvec[0.7]{h_1,h_2,h_3 \\ g_1,g_2,g_3} = (-1)^{4\sum_{\alpha \neq \beta\in\{1,2,3\}}h_\alpha g_\beta} ~,
\ee
which is a set of signs corresponding to a discrete torsion. Notice that $\T$ involves only the $h_\alpha,g_\beta$'s associated with the first three generators, \ie those which act on 16 fermions $\psi^i$'s. From the orbifold point of view, this torsion term corresponds to one among many others that are allowed by modular invariance and captured by alternative choices of $\pm{1\over 2}$ background values. 

$\bullet$ Secondly, there is a phase
\be
\Phi\colvec[0.7]{h_1,\dots,h_4 \\ g_1,\dots,g_4} = e^{-2i\pi\sum_{\alpha=1}^4(h_4g_\alpha+g_4h_\alpha)}~,
\ee
which is somehow a generalization of the sign $(-1)^{4\tilde h\tilde g}$ showing up in \Eq{orb1} in the orbifold case involving only $\tilde G^\f$.

$\bullet$  Finally, we have upgraded the definition~(\ref{ga}) of the  lattice of zero modes of $T^{10-d}$,  
\begin{align}
&\Gamma_{10-d,10-d}\colvec[0.7]{\goh_1,\dots,\goh_d;\goh'_1,\dots,\goh'_d \\ \gog_1,\dots,\gog_d;\gog'_1,\dots,\gog'_d}(G,B)\nonumber \\
&\!\!={\sqrt{\det G}\over (\Im\tau)^{10-d\over2}}\sum_{\substack{\ell_{d},\dots,\ell_9 \\ \tilde k_{d},\dots, \tilde k_9}} \!\!e^{-2i\pi \left(\gog'_{10-K} \ell_K-\goh'_{10-K} \tilde k_K\right)} e^{-{\pi\over \Im\tau}\left[\tilde k_I+\gog_{10-I}+(\ell_I+\goh_{10-I}) \tau\right](G+B)_{IJ}\left[\tilde k_J+\gog_{10-J}+(\ell_J+\goh_{10-J}) \bar\tau\right]\nonumber }\\
&\!\!= e^{-2i\pi \gog_{10-L}\goh'_{10-L}}\sum_{\substack{m_{d},\dots,m_9 \\ \ell_{d},\dots, \ell_9}}e^{-2i\pi \left(\gog_{10-K} m_K+\gog'_{10-K}\ell_K\right)}\,  \bar q^{{1\over 4}P^\text{L}_IG^{-1}_{IJ}P^\text{L}_J}\; q^{{1\over 4}P^\text{R}_IG^{-1}_{IJ}P^\text{R}_J} \, ,
\label{mwshift}
\end{align}
where  we denote in the Hamiltonian form
\be
P^\text{L}_I=m_I+\goh'_{10-I}+(B+G)_{IJ}\, (\ell_J+\goh_{10-J})~ , \quad P^\text{R}_I=m_I+\goh'_{10-I}+(B-G)_{IJ}\, (\ell_J+\goh_{10-J})~.
\ee
In the case at hand, $d=6$ and the $h_\alpha',g_\alpha'$ are not independent as they are expressed in terms of the $h_\beta,g_\beta$,  
\be
\begin{aligned}
\goh'_1&= \goh'_2= \goh'_3= h_4~,\quad &&\goh'_4= h_1+h_2+h_3~ ,\\
\gog'_1&= \gog'_2=\gog'_3= g_4~ ,\quad &&\gog'_4= g_1+g_2+g_3~ .\\
\end{aligned}
\label{shifts}
\ee

To understand the meaning of the general expression of the lattice~(\ref{mwshift}), note that under a double Poisson summation on the Lagrangian expression, \ie on both $\ell_I$ and $\tilde k_I$, $I\in\{d,\dots,9\}$, the roles of $h_{10-I},g_{10-I}$  and $h'_{10-I},g'_{10-I}$ are exchanged. Equivalently, one can perform a T-duality in the Hamiltonian form, which also exchanges the momenta and winding numbers. Hence, the lattice~(\ref{mwshift}) is the suitable one for describing a simultaneous shift action on the coordinates $X^I\equiv X^I_\rL+X^I_\rR$ and  on the T-dual coordinates $\widetilde X^I\equiv X^I_\rL-X^I_\rR$, where $X^I_\rL,X^I_\rR$ are the left- and right-moving parts of the worldsheet coordinates. When the $h'_{10-I},g'_{10-I}$ are not identically vanishing, the orbifold action does not admit a geometric picture and is said (left/right) asymmetric~\cite{asymmorb}. Even if it is a misnomer, it is also referred to as a ``free'' action (even though not geometrical), whose effect is to induce masses in the twisted sectors. That being said, we are now ready to identify from \Eq{shifts} precise definitions of the generators $G_1^{\f\prime},G_2^{\f\prime},G_3^{\f\prime}, \tilde G_4^\f$ in terms of half period shifts of the internal coordinates or T-dual coordinates, 
\be
\begin{aligned}
G_{10-I}^{\f\prime} &= G_{10-I} \otimes \big( X^I\to X^I+\pi \big) \otimes \big( \widetilde X^6\to \widetilde X^6+\pi \big)~,~~\quad  I\in\{7,8,9\}~,\\
\tilde G_{4}^\f &= \tilde G_{4} \otimes \big(  X^6\to  X^6+\pi \big)\otimes \big( \widetilde X^J\to \widetilde X^J+\pi ~, ~ J\in\{7,8,9\}\big)~.\esp
\end{aligned}
\ee

In order to determine the gauge symmetry generated by the fermions $\psi^i$ for generic $4G_{IJ}, 4B_{IJ}$ at the orbifold point, it is enough to analyze the untwisted sector at zero momentum and winding numbers. In that case, we obtain from \Eq{G6}
\begin{align}
 {1\over 2^4} \sum_{g_1,\dots, g_4}&\,  \Gamma_{4,4}\colvec[0.7]{0,\dots,0;0,\dots,0  \\ 
\gog_1,\dots,\gog_4;\gog'_1,\dots,\gog'_4}(4G,4B)\Big|_{\substack{m_6=\dots=m_9=0 \\ \ell_6=\dots= \ell_9=0}}\, {\Gamma^\Spin\colvec[0.7]{0,\dots,0 \\ g_1,\dots,g_4}\over \eta^{16}}\,\Phi \colvec[0.7]{0,\dots,0 \\ g_1,\dots,g_4}\;\T\colvec[0.7]{0,0,0 \\ g_1,g_2,g_3}\nonumber \\
&= {1\over 2^4} \sum_{g_1,\dots, g_4}  {\Gamma^\Spin\colvec[0.7]{0,\dots,0 \\ g_1,\dots,g_4}\over \eta^{16}}=\half\sum_{g_4}\big( O_3O_1 + V_3(-1)^{2g_4}V_1\big)^8 +\text{massive}\\
&=(O_3O_1)^8+\text{massive}~,\nonumber \esp
\end{align}
 where we have used \Eq{g1g2g3} and the decomposition formula and transformations rules of characters listed in the Appendix. Hence, for generic torus moduli, the gauge symmetry generated by the fermions $\psi^i$'s is $\big(SO(3)\times SO(1)\big)^8$, which is of reduced rank 8, as anticipated from the deformations~(\ref{deforma}).  

Before concluding, let us mention that the lattice in \Eq{G6} is exactly that found in \Ref{so3so1}, up to the discrete torsion which was chosen to be trivial in that work, $\T\equiv 1$.$^{\ref{nonsysy}}$ The derivation presented in \Ref{so3so1} was based from the outset on the orbifold point of view, and the phase $\Phi$ along with the asymmetric  orbifold parameters $h'_\alpha,g'_\alpha$, $\alpha\in\{1,\dots, 4\}$, were found on considerations based  on modular invariance only. Hence, the precise form of $\Phi$ as well as the action on the T-dual coordinates are mandatory  by consistency. The advantage of the method based on background deformations presented in this section is that consistency is automatically taken into account by \Eq{block} and ({\ref{blocky}). Moreover, these general expressions  can be used to provide the explicit deformations of the lattices induced by the continuous Wilson-line moduli (see the example of \Eq{WLex}).  Finally, notice that for the $\Z_2^5$ free-orbifold model that leads to the trivial $SO(1)^{32}$ gauge symmetry, we have recovered extra gauge bosons in \Sect{orb1616tw} in a twisted sector by turning the action of one generator into a non-free version. In the present example where the $\Z_2^4$ free-orbifold action yields  $\big(SO(3)\times SO(1)\big)^8$, it is however not possible to play the same game for any generator. Indeed, modular invariance is at the heart of the obstruction, since the action of all generators  must be asymmetric and thus non-trivial on the torus coordinates.


\section{Conclusion}
\label{cl}

String theory models describing non simply-laced gauge symmetry groups have been much less studied and used in the literature. The reasons for this are not fundamental. In many cases, they are instead  consequences of simplifying assumptions that restrict the allowed worldsheet conformal field theories and their associated affine Lie symmetries. For example, in the maximally supersymmetric case, the moduli space of the heterotic string in $d$ spacetime dimensions already admits a multitude of distinct components~\cite{triple}, and it turns out that some of them cannot be described in certain formalisms.

For instance, if one insists on realizing the system of central charge $c=24$ of the bosonic side of the heterotic string in terms of 24 real scalars, one can certainly describe the moduli-space component (for a given spacetime dimension $d\le 9$) that contains the plain $\Spin$ and $E_8\times E_8$ theories. The gauge symmetry described at arbitrary point of this component is then of maximal rank. Of course, these conclusions persist if one replaces the description of some of the 24 bosons by complex fermions in fermionic language.  On the contrary, all components of the moduli space containing models describing gauge group factors  $SO(2n+1)$ cannot be captured in these frameworks. Instead, one is led to consider real fermions with boundary conditions on the worldsheet that do not allow any pairing into complex fermions. The relevant two-dimensional conformal blocks are therefore of  Ising type, with central charge $c=\half$. Note, however, that the existence of non simply-laced gauge symmetry groups in components of the moduli space takes place only at particular loci. Indeed, these groups can be broken into products of non-Abelian unitary, $SO(2n)$ and of course Abelian $U(1)$'s by exploring the Coulomb branches. 

In our work, we have constructed and analyzed in great details two models describing $SO(1)^{32}$ and $\big(SO(3)\times SO(1)\big)^8$ gauge symmetries, respectively in five and six dimensions. Together with their descendants obtained by toroidal compactification, they are representative models of components of the moduli space that exist respectively in dimension $d\le 5$ and $d\le 6$, but not higher~\cite{so3so1,triple}. In the former components (existing for $d\le 5$), the gauge-group ranks are reduced by 16 units, while in the second series of components (existing for $d\le 6$), the ranks  are reduced by 8 units. 

Even though these series of components are somehow associated  with distinct theories, we have seen that the lattices $\Gamma_{10-d,26-d}$ of their respective models can be described in a unified way. 
This approach is  based on the fact that the transformation rules of the lattices under the modular group must be independent of their deformations. This very fact must hold whether the deformations are continuous Wilson lines parametrizing a given component of  the moduli space, or discrete quantities responsible for the switch from one component to the other. The lattices constructed this way can also  be described from a free-orbifold point of view, where the actions of the generators on the gauge degrees of freedom as well as on the internal torus can be identified unambiguously. These actions can be symmetric or asymmetric on the internal space, and with or without discrete torsion. 

Finally, we stress that continuous and discrete deformations in the heterotic framework have counterparts in dual orientifold descriptions in terms of dynamical and rigid D-brane positions. Both points of view can be useful to analyze the fate of the stability at the quantum level of these models in presence of an extra implementation of supersymmetry breaking~\cite{ADLP, ACP-1, so3so1,ACP-2}.


\vspace{0.6cm}
\section*{Acknowledgements} 

\noindent The authors would like to thank Carlo Angelantonj, Emilian Dudas and Gianfranco Pradisi for discussions and useful inputs during the realization of this work. 


\begin{appendix}
\makeatletter
\DeclareRobustCommand{\@seccntformat}[1]{%
  \def\temp@@a{#1}%
  \def\temp@@b{section}%
  \ifx\temp@@a\temp@@b
  \appendixname\ \thesection:\quad%
  \else
  \csname the#1\endcsname\quad%
  \fi
} 
\makeatother

\section*{\boldmath Appendix}
\label{A0}
\renewcommand{\theequation}{A.\arabic{equation}}

In this Appendix, we first review the definitions and $q$-expansions of the affine characters of $SO(n)$. We verify that the low lying modes of the characters are in one-to-one correspondence with the weights of the lowest dimensional representations in each conjugacy class.  Then, we  list formulas satisfied by these characters  and that are  useful for implementing $\Z_2$-orbifold actions. 
 
The affine characters $O_{2n}, V_{2n}, S_{2n},C_{2n}$ of $SO(2n)$~\cite{characters1,characters2,BLT} and   $O_{2n+1}, V_{2n+1}, S_{2n+1}$ of $SO(2n+1)$~\cite{odd_characters} can be expressed in terms of Jacobi theta functions $\vartheta\colvec[0.7]{\gamma\\\delta}(0|\tau)$ with half-integer characteristics $\gamma,\delta$ and Dedekind functions $\eta(\tau)$ as follows, 
\be
\label{defchar}
\begin{aligned}
O_n &= \frac{1}{2\,\eta^{n\over2}}\,\Big( \vartheta\colvec[0.7]{ \overset{}0\\ \overset{}0}^{n\over2} + \vartheta\colvec[0.7]{0\\1/2}^{n\over2} \Big) &&={1\over q^{n\over 48}}\,\Big[1+{n(n-1)\over 2}\,q+\O(q^2)\Big]\, , \\
V_n &= \frac{1}{2\,\eta^{n\over2}}\,\Big( \vartheta\colvec[0.7]{ \overset{}0\\ \overset{}0}^{n\over2} - \vartheta\colvec[0.7]{0\\1/2}^{n\over2} \Big) &&={1\over q^{n\over 48}}\,n\,q^{\half}\big(1+\O(q)\big)\, , \\
S_{2n} &= \frac{1}{2\,\eta^{n}}\,\Big( \vartheta\colvec[0.7]{ \overset{}1/2\\  \overset{}0}^{n} + i^{-n}\,\vartheta\colvec[0.7]{1/2\\1/2}^{n} \Big) &&={1\over q^{2n\over 48}}\,2^n\,q^{n\over 8}\big(1+\O(q)\big)\, , \\
C_{2n} &= \frac{1}{2\,\eta^{n}}\,\Big( \vartheta\colvec[0.7]{ \overset{}1/2\\ \overset{}0}^{n} - i^{-n}\,\vartheta\colvec[0.7]{1/2\\1/2}^{n} \Big) &&={1\over q^{2n\over 48}}\,2^n\,q^{n\over 8}\big(1+\O(q)\big)\, , \\
S_{2n+1}& = \frac{1}{\sqrt{2}\,\eta^{n+\half}}\,\vartheta\colvec[0.7]{ \overset{}1/2\\ \overset{}0}^{2n+1\over2} &&={1\over q^{2n+1\over 48}}\,2^n\,q^{n+\half\over 8}\big(1+\O(q)\big)\, ,
\end{aligned}
\ee
where our notations and conventions for the modular forms are those of Ref.~\cite{review_carlo,OVSC}.\footnote{The dictionary with the conventions of \Ref{Kbook} is given by $\vartheta\colvec[0.7]{\gamma\\\delta}(\nu |\tau)=\theta\colvec[0.7]{2\gamma\\ 2\delta}(-\nu|\tau)= \theta\colvec[0.7]{-2\gamma\\ -2\delta}(\nu|\tau)$.} 


\paragraph{\em \boldmath Characters $O_n$:}

$O_{2n}$ is the character that captures the weights of the singlet and adjoint representations of $SO(2n)$. From the definition of the Jacobi modular forms, we have 
\begin{align}
O_{2n} &= {1\over 2\, q^{n\over 24}}\,\bigg( \prod_{k\ge 1} (1+q^{k-\frac{1}{2}})^{2n} + \prod_{k\ge 1} (1-q^{k-\frac{1}{2}})^{2n} \bigg) \\
&={1\over 2\, q^{n\over 24}}\,\bigg( \Big[(1+q^{\frac{1}{2}(+1)^2})(1+q^{\frac{1}{2}(-1)^2})\Big]^n + \Big[(1-q^{\frac{1}{2}(+1)^2})(1-q^{\frac{1}{2}(-1)^2})\Big]^n + {\cal O}(q^{2}) \bigg)\, .\nonumber 
\end{align}
In the big parentheses, only even powers of $q^\half$ survive. The terms at order $q$ arise by multiplying pairs of $q^{{1\over2}\epsilon_u^2}$,  where $\epsilon_u=\pm 1$, $u\in\{1,...,n\}$.  This pairing can be done by combining an $\epsilon_u$ with $-\epsilon_u$, or by combining $\epsilon_u$ with $\epsilon_v$ for $u<v$. Hence, we obtain 
\be
O_{2n} =  {1\over  q^{n\over 24}}\Big( 1 + nq + \sum_{\epsilon_1,\epsilon_2=\pm 1}q^{\frac{1}{2}[\epsilon_1^2+\epsilon_2^2+0^{n-2}]} + \mbox{permut. $1,2\to u<v\in\{1,\dots,n\}$}+ {\cal O}(q^2) \Big) \, ,
\ee
where $0^{k}$ stands for a sum of $k$ consecutive $0$'s.  In this form, the leading term $q^0$ is associated with the singlet representation of $SO(2n)$, while at order $q$ there are contributions associated with the Cartan subalgebra and the roots of the adjoint representation of $SO(2n)$. Indeed, we recognize in the brackets the roots squared equal to 2 of $SO(2n)$.

For the character $O_{2n+1}$ of $SO(2n+1)$, we have similarly 
\be
\begin{aligned}
O_{2n+1} &= {1\over 2\, q^{n+\half\over 24}}\,\bigg( \prod_{k\ge 1} (1+q^{k-\frac{1}{2}})^{2n+1} + \prod_{k\ge 1} (1-q^{k-\frac{1}{2}})^{2n+1} \bigg) \\
&={1\over 2\, q^{n+\half\over 24}}\,\bigg( \Big[(1+q^{\frac{1}{2}(+1)^2})(1+q^{\frac{1}{2}(-1)^2})\Big]^n(1+q^\half) \\
&\qquad\qquad +\Big[(1-q^{\frac{1}{2}(+1)^2})(1-q^{\frac{1}{2}(-1)^2})\Big]^n(1-q^\half)  + {\cal O}(q^{2}) \bigg)\, .
\end{aligned}
\ee
The contributions at order $q$ contain those associated with the subgroup $SO(2n)$, as well as those obtained by pairing the extra $q^\half$ with all $q^{\half \epsilon_u^2}$'s. In total, we have 
\begin{align}
O_{2n+1} =  {1\over  q^{n+\half\over 24}}\Big( 1 + nq &+ \sum_{\epsilon_1,\epsilon_2=\pm 1}q^{\frac{1}{2}[\epsilon_1^2+\epsilon_2^2+0^{n-2}]} + \mbox{permut. $1,2\to u<v\in\{1,\dots,n\}$}\\
&+\sum_{\epsilon_1=\pm 1}q^{[(\epsilon_1)^2+0^{n-1}]}+ \mbox{permut. $1\to u\in\{1,\dots,n\}$}+ {\cal O}(q^2) \Big) \, ,\nonumber
\end{align}
where the brackets are again interpreted as the roots squared of $SO(2n+1)$~\cite{zuber}. Notice that because $SO(2n+1)$ is not simply laced, the roots not present for $SO(2n)$ have norms squared equal to 1 rather than 2. Thus, there are no  normalization factors $\half$ for them in the exponents of $q$.


\paragraph{\em \boldmath Characters $V_n$:} 

The expansion of the character $V_{2n}$ of $SO(2n)$ is  given by 
\begin{align}
V_{2n} &= {1\over 2\, q^{n\over 24}}\,\bigg( \prod_{k\ge 1} (1+q^{k-\frac{1}{2}})^{2n} - \prod_{k\ge 1} (1-q^{k-\frac{1}{2}})^{2n} \bigg) \\
&={1\over 2\, q^{n\over 24}}\,\bigg( \Big[(1+q^{\frac{1}{2}(+1)^2})(1+q^{\frac{1}{2}(-1)^2})\Big]^n - \Big[(1-q^{\frac{1}{2}(+1)^2})(1-q^{\frac{1}{2}(-1)^2})\Big]^n + {\cal O}\big(q^{3\over 2}\big) \!\bigg)\, , \nonumber 
\end{align}
where only odd powers of $q^\half$ survive in the big parentheses. To make the link with the weights squared of the vectorial representation, the above expression can we rewritten as 
\be
V_{2n} =  {1\over  q^{n\over 24}}\bigg(\sum_{\epsilon_1=\pm 1}q^{\half[\epsilon_1^2+0^{n-1}]} + \mbox{permut. $1\to u\in\{1,\dots,n\}$}\bigg)\big(1+ {\cal O}(q) \big)\, .
\ee

For $SO(2n+1)$ we have in an analogous way 
\be
\begin{aligned}
V_{2n+1} &= {1\over 2\, q^{n+\half\over 24}}\,\bigg( \prod_{k\ge 1} (1+q^{k-\frac{1}{2}})^{2n+1} - \prod_{k\ge 1} (1-q^{k-\frac{1}{2}})^{2n+1} \bigg) \\
&={1\over 2\, q^{n+\half\over 24}}\,\bigg( \Big[(1+q^{\frac{1}{2}(+1)^2})(1+q^{\frac{1}{2}(-1)^2})\Big]^n(1+q^\half) \\
&\qquad\quad~~~ - \Big[(1-q^{\frac{1}{2}(+1)^2})(1-q^{\frac{1}{2}(-1)^2})\Big]^n(1-q^\half)  + {\cal O}\big(q^{3\over  2}\big)\! \bigg)\\
&=  {1\over  q^{n+\half\over 24}}\,\bigg(\sum_{\epsilon_1=\pm 1}q^{\half[\epsilon_1^2+0^{n-1}]} + \mbox{permut. $1\to u\in\{1,\dots,n\}$}+q^\half\bigg)\big(1+ {\cal O}(q) \big)\, .
\end{aligned}
\ee
At leading order in $q$, we recognize the weights squared of the vectorial representation of the subgroup $SO(2n)$ 	along with an extra one to complete the vectorial representation of $SO(2n+1)$. This extra weight is a zero-weight in the same sense that the Cartan subalgebra corresponds to zero-weights of the adjoint representation. In representation theory, it comes from subtracting the extra positive root $(1,0^{n-1})$ of $SO(2n+1)$ (\ie not present for $SO(2n)$) to the highest weight $(1,0^{n-1})$ of the vectorial representation.


\paragraph{\em \boldmath Characters $S_{2n},C_{2n}$ and $S_{2n+1}$:} 

$SO(2n)$ admits spinorial and ``antispinorial'' conjugacy classes and the sum of their respective characters $S_{2n}$ and $C_{2n}$  reads 
\be\label{weightspin}
\begin{aligned}
S_{2n} + C_{2n} &={(2\, q^{1\over 8})^n\over q^{n\over24}}\,\prod_{k\ge 1}(1+q^k)^{2n} \\
&= {1\over q^{n\over 24}}\,\big(q^{\half(+\half)^2}+q^{\half(-\half)^2}\big)^n\big(1+\O(q)\big)	\\
&={1\over q^{n\over24}}\, \bigg( \sum_{\epsilon_1,...,\epsilon_n=\pm1} q^{{1\over2}\left[ (\frac{\epsilon_1}{2})^2 + \cdots + (\frac{\epsilon_n}{2})^2 \right]}\bigg)\big( 1 + {\cal O}(q) \big) \, .
\end{aligned}
\ee
In the last line, we recognize in brackets the weights squared of the two associated spinorial representations of dimensions $2^n/2$. They have opposite chiralities~\cite{Kbook} and correspond to imposing  $\prod_{u=1}^n \epsilon_u =1$ and $\prod_{u=1}^n \epsilon_u =-1$, respectively. 

For $SO(2n+1)$, the expansion of the character of the unique spinorial conjugacy class is
\be
\begin{aligned}
S_{2n+1}  &={(2\, q^{1\over 8})^{n+\half}\over \sqrt{2}\, q^{n+\half\over24}}\,\prod_{k\ge 1}(1+q^k)^{2n+1} \\
&= {q^{1\over 16}\over q^{n\over 24}}\,\big(q^{\half(+\half)^2}+q^{\half(-\half)^2}\big)^n\big(1+\O(q)\big)	\\
&={q^{1\over 16}\over q^{n+\half\over24}}\, \bigg( \sum_{\epsilon_1,...,\epsilon_n=\pm1} q^{{1\over2}\left[ (\frac{\epsilon_1}{2})^2 + \cdots + (\frac{\epsilon_n}{2})^2 \right]}\bigg)\big( 1 + {\cal O}(q) \big) \, .
\end{aligned}
\ee
At leading order, it contains the contributions associated with the $2^n$ weights of the  spinorial representation. The latter cannot be split into spinorial and antispinorial irreducible representations  because of the extra positive root $(1,0^{n-1})$ of $SO(2n+1)$ (as compared to $SO(2n)$). Indeed, when subtracted to the highest weight $(\half,...,\half)$ of the spinorial representation, the number of $+\frac{1}{2}$ becomes odd.


\paragraph{Character decomposition:} 

Because $SO(n)\supset SO(n-p)\times SO(p)$, the representations of $SO(n)$ can be expressed as direct sums of tensor products of representations of $SO(n-p)$ and $SO(p)$. This translates into relations between the characters, which can be derived from the definitions~(\ref{defchar}):
\be
\label{decomp1}
\begin{aligned}
O_n &=O_{n-p}O_p + V_{n-p}V_p ~, \\
 V_n &=O_{n-p}V_p + V_{n-p}O_p ~,\\
S_{2n} &= S_{2n-2p}S_{2p} + C_{2n-2p}C_{2p} \\
&=S_{2n-2p-1}S_{2p+1} + \Delta_{2n-2p-1}\Delta_{2p+1}~,\\
 C_{2n} &=S_{2n-2p}C_{2p} + C_{2n-2p}S_{2p} \\
 &=S_{2n-2p-1}S_{2p+1} - \Delta_{2n-2p-1}\Delta_{2p+1}~,\\
S_{2n+1}&= \big(S_{2n-2p}+C_{2n-2p}\big)S_{2p+1}\\
&= S_{2n+1-2p}\big(S_{2p}+C_{2p}\big)\,,\\
\Delta_{2n+1}&= \big(S_{2n-2p}-C_{2n-2p}\big)\Delta_{2p+1}\\
&= \Delta_{2n+1-2p}\big(S_{2p}-C_{2p}\big)\,.
\end{aligned}
\ee
In the above identities,  we have included vanishing contributions 
\be
\Delta_{2n+1} = \frac{1}{\sqrt{2}\, \eta^{2n+1\over2}}\, \vartheta\colvec[0.7]{1/2\\1/2}^{2n+1\over2} =0~ ,
\label{delta}
\ee 
which have no meaning from the point of view of representation theory and can be omitted, except in the course of a derivation of twisted-sector spectra obtained by flipping boundary conditions along the cycle $[0,1]$ of the genus-one worldsheet (see \Sect{orb1616tw}).


\paragraph{\boldmath Projection onto $\pmb{\Z}_2$-invariant weights:} 

The implementation of a $\Z_2$-orbifold action generated by $G$ on a weight lattice requires a projection on $G$-invariant weights. To this end, one must derive how the characters transform as the generator $G$ is inserted in the trace $\tr q^{L_0-1}\to\tr (Gq^{L_0-1})$. Alternatively, one can flip the boundary conditions along the cycle $[0,\tau]$ of the fermions $\psi^i$, $i\in\{1,\dots,2n\}$, which amounts to shifting the characteristics of the Jacobi modular forms as follows, $\colvec[0.7]{\gamma\\\delta}\to \colvec[0.7]{\gamma\\\delta+\half}$. The transformation rules are then 
\be
\label{Gtr}
\begin{aligned}
(O_n,V_n)&~~ \longrightarrow ~~ (O_n,-V_n)~,\\
(S_{2n},C_{2n})&~~ \longrightarrow~~ i^n(S_{2n},-C_{2n})~,\\
(S_{2n+1},\Delta_{2n+1})&~~ \longrightarrow~~ i^{n+\half}(\Delta_{2n+1},S_{2n+1})~.
\end{aligned}
\ee
Notice that applied twice, they give the initial characters back up to phases because $\vartheta\colvec[0.7]{\gamma\\\delta+\half+\half}= e^{2i\pi \gamma}\vartheta\colvec[0.7]{\gamma\\\delta}$. 


\paragraph{\boldmath $\pmb{\Z}_2$-twisted sector:} 

The implementation of a $\Z_2$-orbifold action on a lattice also implies the existence of a twisted sector. The latter is obtained from the untwisted sector by changing the boundary conditions of the $\psi^i$, $i\in\{1,\dots,2n\}$, along the string \ie the worldsheet cycle $[0,1]$. The characteristics of the Jacobi modular forms are therefore modified as $\colvec[0.7]{\gamma\\\delta}\to \colvec[0.7]{\gamma+\half\\\delta}$. In terms of characters, this corresponds to changing 
\be
\label{Gtwist}
\begin{aligned}
(O_{2n},V_{2n})&~~ \longrightarrow ~~ \left({1+i^n\over 2}\,S_{2n}+{1-i^n\over 2}\,C_{2n}\,,{1-i^n\over 2}\,S_{2n}+{1+i^n\over 2}\,C_{2n}\right)\,,\espD\\
(S_{2n},C_{2n})&~~ \longrightarrow ~~ \left({1+i^{-n}\over 2}\,O_{2n}+{1-i^{-n}\over 2}\,V_{2n}\,,{1-i^{-n}\over 2}\,O_{2n}+{1+i^{-n}\over 2}\,V_{2n}\right)\,,\\
(O_{2n+1},V_{2n+1})&~~ \longrightarrow ~~\left({S_{2n+1}+i^{n+\half}\,\Delta_{2n+1}\over \sqrt{2}}\, , {S_{2n+1}-i^{n+\half}\,\Delta_{2n+1}\over \sqrt{2}}\right)\, , \\
(S_{2n+1},\Delta_{2n+1})&~~ \longrightarrow ~~\left({O_{2n+1}+V_{2n+1}\over \sqrt{2}}\, , i^{-n-\half}{O_{2n+1}-V_{2n+1}\over \sqrt{2}}\right)\, .
\end{aligned}
\ee
Notice that applying twice these transformations gives the initial characters back because $\vartheta\colvec[0.7]{\gamma+\half+\half\\\overset{}\delta}= \vartheta\colvec[0.7]{\gamma\\\delta}$. 

\end{appendix}


\bibliographystyle{unsrt}

\end{document}